\documentclass{iopart}
\usepackage{graphicx}
\usepackage{iopams}

\newcommand{\lya}{Ly$\alpha$~}
\newcommand{\Lya}{Ly$\alpha$~}
\newcommand{\HI}{\rm H\,I~}
\newcommand{\zr}{z_{\rm reion}}
\newcommand{\beq}{\begin{equation}}
\newcommand{\beqa}{\begin{eqnarray}}
\newcommand{\eeq}{\end{equation}}
\newcommand{\eeqa}{\end{eqnarray}}
\newcommand{\Omm}{{\Omega_m}}

\newcommand{\la}{{\lesssim}}

\def\Omr{{\Omega_r}}
\def\Omk{{\Omega_k}}
\def\Oml{{\Omega_{\Lambda}}}

\begin{document}

\review{The Physics and Early History of the Intergalactic Medium}

\author{Rennan Barkana$^1$ and Abraham Loeb$^2$}

\address{$^1$ School of Physics and Astronomy, The Raymond and Beverly 
Sackler Faculty of Exact Sciences, Tel Aviv University, Tel Aviv
69978, ISRAEL}

\address{$^2$ Astronomy Department, Harvard University, 60 Garden
Street, Cambridge, MA 02138, USA}

\eads{\mailto{barkana@wise.tau.ac.il}, \mailto{aloeb@cfa.harvard.edu}}


\begin{abstract}

The intergalactic medium -- the cosmic gas that fills the great spaces
between the galaxies -- is affected by processes ranging from quantum
fluctuations in the very early universe to radiative emission from
newly-formed stars. This gives the intergalactic medium a dual role as
a powerful probe both of fundamental physics and of astrophysics. The
heading of fundamental physics includes conditions in the very early
universe and cosmological parameters that determine the age of the
universe and its matter content. The astrophysics refers to chapters
of the long cosmic history of stars and galaxies that are being
revealed through the effects of stellar feedback on the cosmic
gas. This review describes the physics of the intergalactic medium,
focusing on recent theoretical and observational developments in
understanding early cosmic history. In particular, the earliest
generation of stars is thought to have transformed the universe from
darkness to light and to have had an enormous impact on the
intergalactic medium. Half a million years after the big bang the
universe was filled with atomic hydrogen. As gravity pulled gas clouds
together, the first stars ignited and their radiation turned the
surrounding atoms back into free electrons and ions. From the observed
spectral absorption signatures of the gas between us and distant
sources, we know that the process of reionization pervaded most of
space a billion years after the big bang, so that only a small
fraction of the primordial hydrogen atoms remained between
galaxies. Knowing exactly when and how the reionization process
happened is a primary goal of cosmologists, because this would tell us
when the early stars and black holes formed and in what kinds of
galaxies. The distribution and clustering of these galaxies is
particularly interesting since it is driven by primordial density
fluctuations in the dark matter.

Cosmic reionization is beginning to be understood with the help of
theoretical models and computer simulations. Numerical simulations of
reionization are computationally challenging, as they require radiative
transfer across large cosmological volumes as well as sufficiently high
resolution to identify the sources of the ionizing radiation in the infant
universe. Rapid progress in our understanding is expected with additional
observational input. A wide variety of instruments currently under design
-- including large-aperture infrared telescopes on the ground or in space
({\it JWST}\/), and low-frequency radio telescope arrays for the detection
of red shifted 21-cm radiation -- will probe the first sources of light
during an epoch in cosmic history that has been largely unexplored so
far. The new observations and the challenges for theoretical models and
numerical simulations will motivate intense work in this field over the
coming decade.

\end{abstract}


\maketitle

\section{Introduction}

\subsection{The cosmic history of the intergalactic medium}

\begin{figure}
\centering
\includegraphics[width=5in]{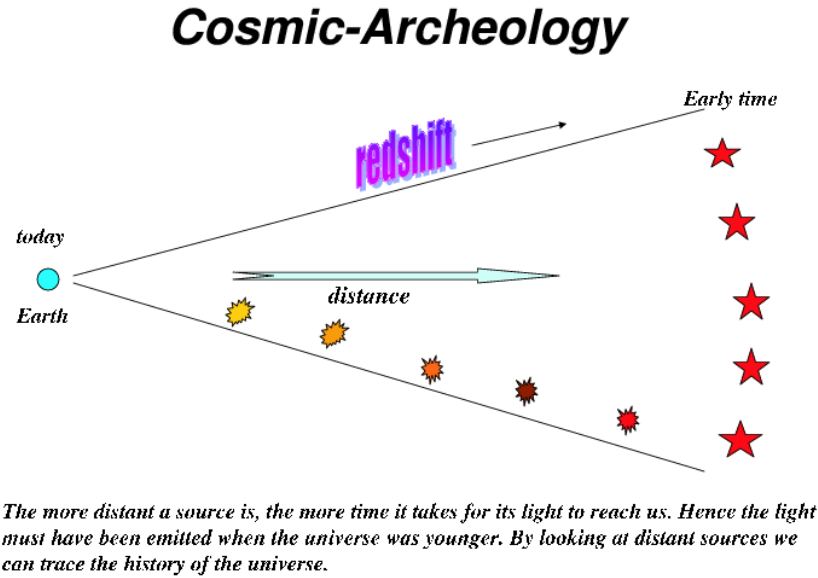}
\caption{Cosmology is like archeology. The deeper one looks, the older is
the layer that is revealed, owing to the finite propagation speed of light
(illustration from Loeb 2006).}
\label{fig:z}
\end{figure}

When we look at our image reflected off a mirror at a distance of 1
meter, we see the way we looked 6.7 nanoseconds ago, the light travel
time to the mirror and back. If the mirror is spaced $10^{19}~{\rm cm}
\simeq 3~$pc away, we will see the way we looked twenty one years
ago. Light propagates at a finite speed, and so by observing distant
regions, we are able to see what the Universe looked like in the past,
a light travel time ago (Figure~\ref{fig:z}). The statistical
homogeneity of the Universe on large scales guarantees that what we
see far away is a fair statistical representation of the conditions
that were present in in our region of the Universe a long time ago.

\begin{figure}
\centering
\includegraphics[width=5in]{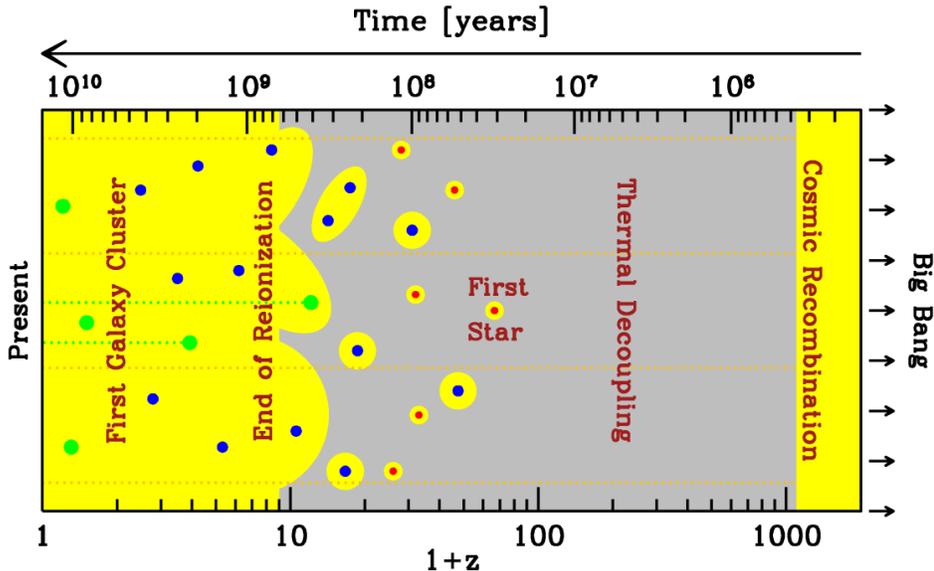}
\caption{Overview of cosmic history, with the age of the universe shown on
the top axis and the corresponding redshift on the bottom axis. Yellow
represents regions where the hydrogen is ionized, and gray, neutral
regions. Stars form in galaxies located within dark matter concentrations
whose typical mass grows with time, starting with $\sim 10^5 M_{\odot}$
(red circles) for the host of the first star, rising to $10^7$--$10^9
M_{\odot}$ (blue circles) for the sources of reionization, and reaching
$\sim 10^{12} M_{\odot}$ (green circles) for present-day galaxies like our
own Milky Way. Astronomers probe the evolution of the cosmic gas using the
absorption of background light (dotted lines) by atomic hydrogen along the
line of sight. The classical technique uses absorption by the
Lyman-$\alpha$ resonance of hydrogen of the light from bright quasars
located within massive galaxies, while a new type of astronomical
observation will use the 21-cm line of hydrogen with the cosmic microwave
background as the background source (illustration from Barkana
2006a).}
\label{fig:history}
\end{figure}

This fortunate situation makes cosmology an empirical science. We do
not need to guess how the Universe evolved. Using telescopes we can
simply see how it appeared at earlier cosmic times. In principle, this
allows the entire 13.7 billion year cosmic history of our universe to
be reconstructed by surveying the galaxies and other sources of light
to large distances (Figure~\ref{fig:history}). Since a greater
distance means a fainter flux from a source of a fixed luminosity, the
observation of the earliest sources of light requires the development
of sensitive instruments and poses challenges to observers.

To measure distance, astronomers use the characteristic emission patterns
of hydrogen and other chemical elements in the spectrum of each galaxy to
measure its cosmological redshift $z$. As the universe expands, photon
wavelengths get stretched as well, so that the spectrum we observe today is
shifted from the emitted one by a factor of $(1+z)$ in wavelength. This
then implies that the universe has expanded by a factor of $(1+z)$ in
linear dimension since that time, and cosmologists can calculate the
corresponding distance and cosmic age for the source galaxy. Large
telescopes have allowed astronomers to observe faint galaxies that are so
far away that we see them more than ten billion years in the past. Thus, we
know directly that galaxies were in existence as early as 850 million years
after the Big Bang, at a redshift of $z \sim 6.5$ (Hu \etal 2002, White
\etal 2003, Iye \etal 2006).

\begin{figure}
\centering
\includegraphics[width=5in]{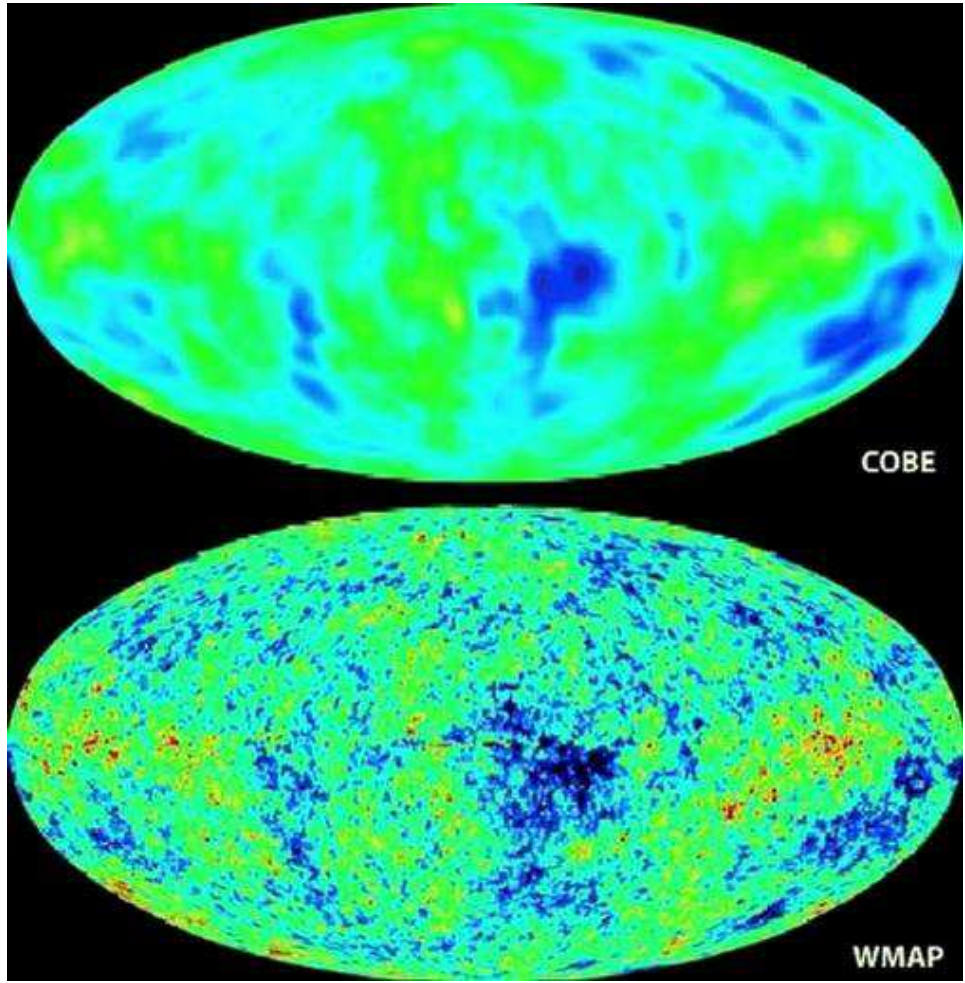}
\caption{Images of the Universe shortly after it became transparent, taken
by the {\it COBE}\/ and {\it WMAP}\/ satellites (see
http://map.gsfc.nasa.gov/ for details). The slight density inhomogeneities
in the otherwise uniform Universe imprinted a map of hot and cold spots
(shown here as different colors) in the CMB that is observed today. The
existence of these anisotropies was predicted three decades before the
technology for taking these images became available, in a number of
theoretical papers including Sachs \& Wolfe (1967), Rees \& Sciama (1968),
Silk (1968), Sunyaev \& Zeldovich (1970), and Peebles \& Yu (1970).}
\label{fig:CMB}
\end{figure}

We can in principle image the Universe only if it is transparent. Earlier
than $400\,000$ years after the big bang, the cosmic gas was ionized and
the Universe was opaque to Thomson scattering by the free electrons in the
dense plasma. Thus, telescopes cannot be used to image the infant Universe
at earlier times (or redshifts $\gtrsim 10^3$). The earliest possible image
of the Universe was recorded by the COBE and WMAP satellites (Bennett \etal
1996; Spergel \etal 2006), which recorded the temperature distribution of
the cosmic microwave background (CMB) on the sky (Figure~\ref{fig:CMB}).

The CMB, the relic radiation from the fiery beginning of the universe, is
indeed another major probe of observational cosmology. The universe cools
as it expands, so it was initially far denser and hotter than it is
today. For hundreds of thousands of years the cosmic gas consisted of a
plasma of protons, electrons, and a slight mix of light nuclei, sustained
by the intense thermal motion of these particles. Just like the plasma in
our own Sun, the ancient cosmic plasma emitted and scattered a strong field
of visible and ultraviolet photons. As mentioned above, about $400\,000$
years after the Big Bang the temperature of the universe dipped for the
first time below a few thousand degrees Kelvin. The protons and electrons
were now moving slowly enough that they could attract each other and form
hydrogen atoms, in a process known as cosmic recombination. With the
scattering of the energetic photons now much reduced, the photons continued
traveling in straight lines, mostly undisturbed except that cosmic
expansion has redshifted them into the microwave regime. The emission
temperature of the observed spectrum of these CMB photons is the same in
all directions to one part in $100\,000$ (Figure~\ref{fig:CMB}), which
reveals that conditions were extremely uniform in the early universe.

It was just before the moment of cosmic recombination (when matter
started to dominate in energy density over radiation) that gravity
entered the scene. Since that time, gravity has amplified the tiny
fluctuations in temperature and density observed in the CMB data
(Spergel \etal 2006). Regions that started out slightly denser than
average began to contract because the gravitational forces were also
slightly stronger than average in these regions. Eventually, after
hundreds of millions of years of contraction, galaxies and the stars
within them were able to form. This process, however, would have taken
too long to explain the abundance of galaxies today, if it involved
only the observed cosmic gas. Instead, gravity is strongly enhanced by
the presence of dark matter -- an unknown substance that makes up the
vast majority (84\%) of the cosmic density of matter. The motion of
stars and gas around the centers of nearby galaxies indicates that
each is surrounded by an extended mass of dark matter, and so
dynamically-relaxed dark matter concentrations are generally referred
to as ``halos''.

According to the standard cosmological model, the dark matter is cold
(abbreviated as CDM), i.e., it behaves as a collection of
collisionless particles that started out at matter domination with
negligible thermal velocities and have evolved exclusively under
gravitational forces. The model explains how both individual galaxies
and the large-scale patterns in their distribution originated from the
small initial density fluctuations. On the largest scales,
observations of the present galaxy distribution have indeed found the
same statistical patterns as seen in the CMB, enhanced as expected by
billions of years of gravitational evolution (Eisenstein \etal 2005;
Cole \etal 2005). On smaller scales, the model describes how regions
that were denser than average collapsed due to their enhanced gravity
and eventually formed gravitationally-bound halos, first on small
spatial scales and later on larger ones. In this hierarchical model of
galaxy formation, the small galaxies formed first and then merged or
accreted gas to form larger galaxies. At each snapshot of this cosmic
evolution, the abundance of collapsed halos, whose masses are
dominated by dark matter, can be computed from the initial conditions
using numerical simulations. The common understanding of galaxy
formation is based on the notion that stars formed out of the gas that
cooled and subsequently condensed to high densities in the cores of
some of these halos.

Gravity thus explains how some gas is pulled into the deep potential wells
within dark matter halos and forms the galaxies. One might naively expect
that the gas outside halos would remain mostly undisturbed. However,
observations show that it has not remained neutral (i.e., in atomic form)
until the present. To learn about diffuse gas pervading the space outside
and between galaxies [referred to as the intergalactic medium, or IGM
(Field 1972)], astronomers study its absorption signatures in the spectra
of distant quasars, the brightest long-lived astronomical objects. Quasars'
great luminosities are believed to be powered by gas accretion onto black
holes weighing up to a few billion times the mass of the Sun that are
situated in the dense centers of massive galaxies. As the surrounding gas
spirals in toward the black hole sink, its excess rotation yields viscous
dissipation of heat that makes the gas glow brightly into space, creating a
luminous source visible from afar.

The Lyman-alpha (Ly$\alpha$) resonance line of hydrogen at a
wavelength of 1216 \AA\ has been widely used to trace hydrogen gas
through its absorption of quasar light (Gunn \& Peterson 1965). The
expansion of the universe gives this tool an important advantage
common to all spectral absorption probes. Since the wavelength of
every photon is stretched as the universe expands, the rest-frame
absorption at 1216 \AA\ by a gas element at redshift $z$ is observed
today at a wavelength of 1216$\, (1+z)$ \AA. The absorptions of the
different gas elements along the line of sight are therefore
distributed over a broad range of wavelengths, making it possible to
measure the distribution of intergalactic hydrogen.

Ly$\alpha$ absorption shows that the IGM has been a hot plasma at
least from a cosmic age of 850 million years ($z \sim 6.5$) until
today (White \etal 2003). Thus, the hydrogen must have been ionized
for a second time after it became neutral at cosmic
recombination. Radiation from the first generations of stars is a
plausible source for the ionizing photons that transformed the IGM.

Absorption at the Ly$\alpha$ resonance is so strong that it becomes
difficult to use as observations approach the reionization epoch where
the density of neutral hydrogen becomes high (White \etal 2003). As
described below, cosmologists believe that a different method, termed
``21-cm cosmology'', will allow us to measure how the reionization
process developed over time and to test theoretical predictions of the
properties of the earliest galaxies (Barkana \& Loeb 2001; Loeb 2006).

\subsection{The expanding universe}

The modern physical description of the Universe as a whole can be
traced back to Einstein, who argued theoretically for the so-called
``cosmological principle'': that the distribution of matter and energy
must be homogeneous and isotropic on the largest scales. Today
isotropy is well established (see the review by Wu, Lahav, \& Rees
1999) for the distribution of faint radio sources, optically-selected
galaxies, the X-ray background, and most importantly the cosmic
microwave background (hereafter, CMB; see, e.g., Bennett \etal
1996). The constraints on homogeneity are less strict, but a
cosmological model in which the Universe is isotropic but
significantly inhomogeneous in spherical shells around our special
location, is also excluded (Goodman \etal 1995).

Cosmological solutions of General Relativity predict the evolution of
the cosmic scale factor $a(t)$, defined so that the physical
separation between observers at rest (with respect to the general
expansion) increases with time in proportion to $a(t)$. A given
observer sees a nearby observer at physical distance $D$ receding at
the Hubble velocity $H(t)D$, where the Hubble constant at time $t$ is
$H(t)=d\,a(t)/dt$. Light emitted by a source at time $t$ with an
emitted wavelength $\lambda_{\rm emit}$ is observed at $t=0$ with a
redshifted wavelength $\lambda_{\rm obs}$; the source redshift $z$,
defined from $\lambda_{\rm obs} = \lambda_{\rm emit} \times (1+z)$, is
related to the cosmic scale factor at the time of emission by
$z=1/a(t)-1$, where we set $a(t=0) \equiv 1$ for convenience.

The Einstein field equations of General Relativity yield the Friedmann
equation (e.g., Weinberg 1972; Kolb \& Turner 1990) \beq
H^2(t)=\frac{8 \pi G}{3}\rho-\frac{k}{a^2}\ ,\eeq which relates the
expansion of the Universe to its matter-energy content. The constant
$k$ determines the geometry of the universe; it is positive in a
closed universe, zero in a flat universe, and negative in an open
universe. For each component of the energy density $\rho$, with an
equation of state $p=p(\rho)$, the density $\rho$ varies with $a(t)$
according to the equation of energy conservation \beq d (\rho R^3)=-p
d(R^3)\ . \eeq With the critical density \beq \rho_C(t) \equiv \frac{3
H^2(t)}{8 \pi G} \eeq defined as the density needed for $k=0$, we
define the ratio of the total density to the critical density as \beq
\Omega \equiv \frac{\rho}{\rho_C}\ . \eeq With $\Omm$, $\Oml$, and
$\Omr$ denoting the present contributions to $\Omega$ from matter
(including cold dark matter as well as a contribution $\Omega_b$ from
baryons), vacuum density (cosmological constant), and radiation,
respectively, the Friedmann equation becomes \beq \frac{H(t)}{H_0}=
\left[ \frac{\Omm} {a^3}+ \Oml+ \frac{\Omr}{a^4}+
\frac{\Omk}{a^2}\right]\ , \eeq where we define $H_0$ and
$\Omega_0=\Omm+\Oml+\Omr$ to be the present values of $H$ and
$\Omega$, respectively, and we let \beq \Omk \equiv
-\frac{k}{H_0^2}=1-\Omega_m. \eeq In the particularly simple
Einstein-de Sitter model ($\Omm=1$, $\Oml=\Omr=\Omk=0$), the scale
factor varies as $a(t) \propto t^{2/3}$. Even models with non-zero
$\Oml$ or $\Omk$ approach the Einstein-de Sitter behavior at high
redshift, i.e.\ when $(1+z) \gg |\Omm^{-1}-1|$ (as long as $\Omr$ can
be neglected). In this high-$z$ regime the age of the Universe is
\begin{equation}
t\approx {2\over 3 H_0 \sqrt{\Omega_m}} \left(1+z\right)^{-3/2}\ .
\end{equation}

Recent observations confine the standard set of cosmological
parameters to a relatively narrow range. In particular, we seem to
live in a $\Lambda$CDM cosmology (with $\Omk$ so small that it is
usually assumed to equal zero) with an approximately scale-invariant
primordial power spectrum of density fluctuations, i.e., $n \approx 1$
where the initial power spectrum is $P(k) \propto k^n$ in terms of the
wavenumber $k$. Also, the Hubble constant today is written as
$H_0=100h \mbox{ km s}^{-1}\mbox{Mpc}^{-1}$ in terms of $h$, and the
overall normalization of the power spectrum is specified in terms of
$\sigma_8$, the root-mean-square amplitude of mass fluctuations in
spheres of radius $8\ h^{-1}$ Mpc. For example, the best-fit
cosmological parameters matching the three-year WMAP data together
with large-scale gravitational lensing observations (Spergel \etal
2006) are $\sigma_8=0.826$, $n=0.953$, $h=0.687$, $\Omega_m=0.299$,
$\Omega_\Lambda=0.701$ and $\Omega_b=0.0478$. A different cosmological
parameter set, also based on the CMB data together with other
large-scale structure measurements is (Viel \etal 2006):
$\sigma_8=0.785$, $n=0.957$, $h=0.723$ $\Omega_m=0.253$,
$\Omega_\Lambda=0.747$, and $\Omega_b=0.0425$. The difference between
these two parameter sets roughly represents current $1-\sigma$
parameter uncertainties.

\section{Atomic physics of the intergalactic medium}

\subsection{Radiative absorption and scattering}

\label{sec:absorb}

Resonant Ly$\alpha$ absorption has thus far proved to be the best
probe of the state of the IGM. The optical depth to absorption by a
uniform intergalactic medium is (Gunn \& Peterson 1965)
\beqa
\tau_{s}&=&{\pi e^2 f_\alpha \lambda_\alpha n_{\HI}(z) \over
m_e cH(z)} \label{G-P} \\ \nonumber
&\approx& 6.45\times 10^5 x_{\HI} \left({\Omega_bh\over
0.0315}\right)\left({\Omega_m\over 0.3}\right)^{-1/2}
\left({1+z\over 10}\right)^{3/2}\ , 
\eeqa
where $H\approx 100h~{\rm km~s^{-1}~Mpc^{-1}} \Omega_m^{1/2}
(1+z)^{3/2}$ is the Hubble parameter at redshift $z$;
$f_\alpha=0.4162$ and $\lambda_\alpha=1216$\AA\, are the oscillator
strength and the wavelength of the Ly$\alpha$ transition; $n_{\HI}(z)$
is the neutral hydrogen density at $z$ (assuming primordial
abundances); $\Omega_m$ and $\Omega_b$ are the present-day density
parameters of all matter and of baryons, respectively; and $x_{\HI}$
is the average fraction of neutral hydrogen. In the second equality we
have implicitly considered high redshifts.

\begin{figure}
\centering
\includegraphics[width=5in]{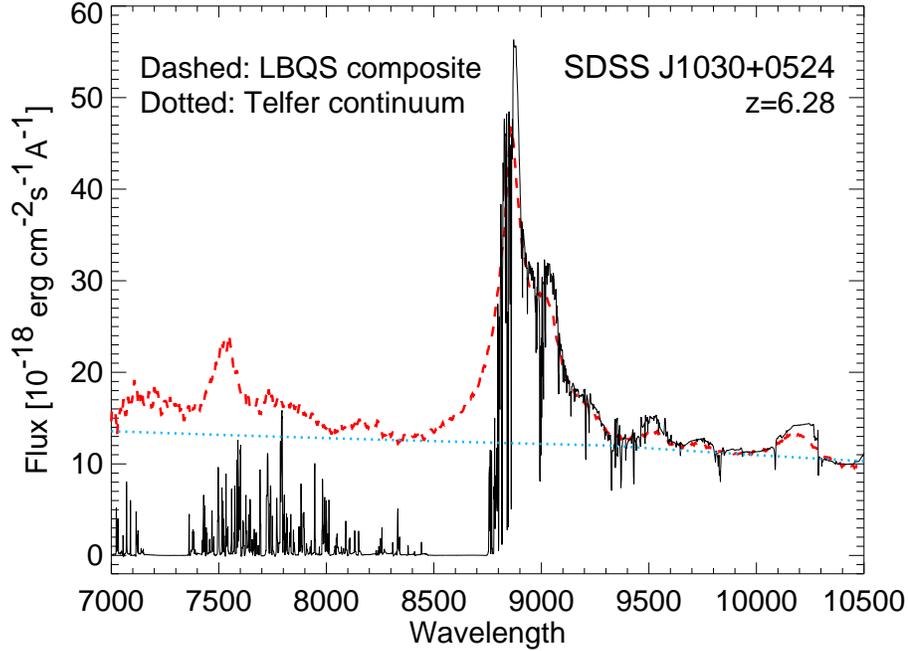}
\caption{Using Ly$\alpha$ absorption in quasar spectra to probe the 
ionization state of the IGM. This figure from White \etal (2003) shows
the observed spectrum of a $z = 6.28$ quasar (solid curve), and the
expected unabsorbed emission (dashed curve), based on an average over
many quasars seen at lower redshifts. The unabsorbed emission is a
sum of smooth emission (the "continuum", dotted curve) plus emission
features from atomic resonances ("emission lines").}
\label{fig:white}
\end{figure}

Ly$\alpha$ absorption is thus highly sensitive to the presence of even
trace amounts of neutral hydrogen. The lack of full absorption in quasar
spectra then implies that the IGM has been very highly ionized during much
of the history of the universe, from the present out to high redshift. At
redshifts approaching six, however, the optical depth increases, and the
observed absorption becomes very strong.  An example of this is shown in
Figure~\ref{fig:white}, taken from White \etal (2003), where an observed
quasar spectrum is compared to the unabsorbed expectation for the same
quasar. The prominent Ly$\alpha$ emission line, which is produced by
radiating hot gas near the quasar itself, is centered at a wavelength of
8850\AA, which for the redshift (6.28) of this quasar corresponds to a
rest-frame 1216\AA. Above this wavelength, the original emitted quasar
spectrum is seen, since photons emitted with wavelengths higher than
1216\AA\ redshift to higher wavelengths during their journey toward us and
never encounter resonance lines of hydrogen atoms. Shorter-wavelength
photons, however, redshift until they hit the local 1216\AA\ and are then
absorbed by any existing hydrogen atoms. The difference between the
unabsorbed expectation and the actual observed spectrum can be used to
measure the amount of absorption, and thus to infer the atomic hydrogen
density. In this particular quasar, this difference is very large (i.e.,
the observed flux is near zero) just to the blue of the Ly$\alpha$ emission
line.

\begin{figure}
\centering
\includegraphics[width=5in]{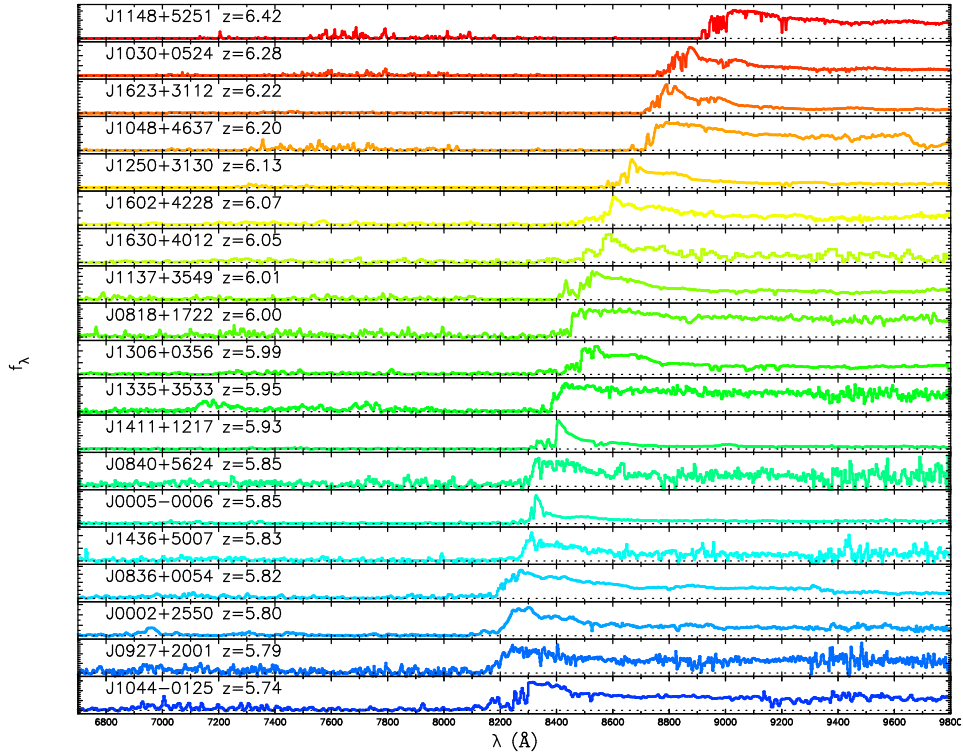}
\caption{Spectra of 19 quasars with redshifts $5.74<z<6.42$ from the 
{\it Sloan Digital Sky Survey}, taken from Fan \etal (2005). For some
of the highest-redshift quasars, the spectrum shows no transmitted
flux shortward of the Ly$\alpha$ wavelength at the quasar redshift
(the so-called ``Gunn-Peterson trough''), indicating a non-negligible
neutral fraction in the IGM.}
\label{fig:19qsos}
\end{figure}

Several quasars beyond $z\sim6.1$ show in their spectra such a
Gunn-Peterson trough, a blank spectral region at wavelengths shorter
than \lya at the quasar redshift (Figure~\ref{fig:19qsos}). The
detection of Gunn-Peterson troughs indicates a rapid change (Fan \etal
2002, White \etal 2003, Fan \etal 2006b) in the neutral content of the
IGM at $z\sim6$, and hence a rapid change in the intensity of the
background ionizing flux. However, even a small atomic hydrogen
fraction of $\sim 10^{-3}$ would still produce nearly complete \lya
absorption. The lower absorption efficiencies of higher Lyman
transitions allow much stronger constraints to be derived from
observations of Ly$\beta$ and Ly$\gamma$ absorption (Fan \etal 2006a).

While only resonant \lya absorption is important at moderate
redshifts, the damping wing of the \lya line plays a significant role
when neutral fractions of order unity are considered at $z \gtrsim 6$.
The scattering cross-section of the \lya resonance line by neutral
hydrogen is given by (Section~23 of Peebles 1993)
\begin{equation}
\sigma_\alpha(\nu) = {3 \lambda_\alpha^2 \Lambda_\alpha^2 \over 8\pi}
{(\nu/\nu_\alpha)^4\over
4\pi^2(\nu-\nu_\alpha)^2+(\Lambda_\alpha^2/4)(\nu/\nu_\alpha)^6}\ ,
\label{eq:sig}
\end{equation}
where $\Lambda_\alpha=(8\pi^2 e^2
f_\alpha/3m_ec\lambda_\alpha^2)=6.25\times 10^8~{\rm s^{-1}}$ is the
\lya ($2p\rightarrow 1s$) decay rate, $f_\alpha=0.4162$ is the
oscillator strength, and $\lambda_\alpha=1216$\AA\, and
$\nu_\alpha=(c/\lambda_\alpha)=2.47\times 10^{15}~{\rm Hz}$ are the
wavelength and frequency of the \lya line. The term in the numerator
is responsible for the classical Rayleigh scattering.

While reionization is a quite inhomogeneous process (as we discuss
below), we consider here a simple illustrative case of instantaneous
reionization. Consider a source at a redshift $z_s$ beyond the
redshift of reionization, $\zr$, and the corresponding scattering
optical depth of a uniform, neutral IGM of hydrogen density $n_{\rm
H,0}(1+z)^3$ between the source and the reionization redshift. The
optical depth is a function of the observed wavelength $\lambda_{\rm
obs}$,
\begin{equation}
\tau(\lambda_{\rm obs})=\int_{\zr}^{z_s} dz\, {cdt\over dz}\, n_{\rm
H,0} (1+z)^3 \sigma_\alpha\left[\nu_{\rm obs}(1+z)\right]\ ,
\end{equation}
where $\nu_{\rm obs}=c/\lambda_{\rm obs}$ and
\beqa
{dt\over dz} = &
\left[(1+z)H(z)\right]^{-1}=H_0^{-1} \times \nonumber \\
&\left[\Omega_m(1+z)^5+\Omega_\Lambda(1+z)^2+
(1-\Omega_m-\Omega_\Lambda)(1+z)^4\right]^{-1/2}\ .
\eeqa

At wavelengths longer than \lya at the source, the optical depth
obtains a small value; these photons redshift away from the line
center along its red wing and never resonate with the line core on
their way to the observer.  Considering only the regime in which
$\vert\nu-\nu_\alpha\vert \gg \Lambda_\alpha$, we may ignore the
second term in the denominator of equation~(\ref{eq:sig}). This leads
to an analytical result for the red damping wing of the Gunn-Peterson
trough (Miralda-Escud\'e 1998).

At wavelengths shorter than 912\AA, the photons are absorbed when they
photoionize atoms of hydrogen or helium. The detailed absorption
cross-sections are summarized in Barkana \& Loeb (2001). For rough
estimates, the average photoionization cross-section for a mixture of
hydrogen and helium with cosmic abundances can be approximated in the
range of $54<h\nu \lesssim 10^3$ eV as $\sigma_{bf}\approx \sigma_0
(\nu/\nu_{\rm H,0})^{-3}$, where $\sigma_0\approx 6\times
10^{-17}~{\rm cm^2}$ (Miralda-Escud\'e 2000).  Before reionization, in
a universe filled with atoms, this yields a very high optical depth
for the absorption of ionizing photons with tens of eV's in
energy. The bound-free optical depth only becomes of order unity in
the extreme ultraviolet (UV) to soft X-rays, around $h\nu \sim 0.1$
keV, a regime which is unfortunately difficult to observe due to
Galactic absorption (Miralda-Escud\'e 2000).

\subsection{The spin temperature of the 21-cm transition of hydrogen}

\label{sec:21cmAtomic}

The ground state of hydrogen exhibits hyperfine splitting involving
the spins of the proton and the electron. The state with parallel
spins (the triplet state) has a slightly higher energy than the state
with anti-parallel spins (the singlet state). The 21-cm line
associated with the spin-flip transition from the triplet to the
singlet state is often used to detect neutral hydrogen in the local
universe. At high redshift, the occurrence of a neutral
pre-reionization IGM offers the prospect of detecting the first
sources of radiation and probing the reionization era by mapping the
21-cm emission from neutral regions. While its energy density is
estimated to be only a $1\%$ correction to that of the CMB, the
redshifted 21-cm emission should display angular structure as well as
frequency structure due to inhomogeneities in the gas density field
(Hogan \& Rees 1979; Scott \& Rees 1990), hydrogen ionized fraction,
and spin temperature (Madau \etal 1997). Indeed, a full mapping of the
distribution of H~I as a function of redshift is possible in
principle.

The basic physics of the hydrogen spin transition is determined as
follows (for a more detailed treatment, see Madau et al.\ 1997 and
Furlanetto \etal 2006). The ground-state hyperfine levels of hydrogen
tend to thermalize with the CMB background, making the IGM
unobservable. If other processes shift the hyperfine level populations
away from thermal equilibrium, then the gas becomes observable against
the CMB in emission or in absorption. The relative occupancy of the
spin levels is usually described in terms of the hydrogen spin
temperature $T_S$, defined by \beq
\frac{n_1}{n_0}=3\, \exp\left\{-\frac{T_*}{T_S}\right\}\ , \eeq where $n_0$
and $n_1$ refer respectively to the singlet and triplet hyperfine levels in
the atomic ground state ($n=1$), and $T_*=0.068$ K is defined by $k_B
T_*=E_{21}$, where the energy of the 21 cm transition is $E_{21}=5.9 \times
10^{-6}$ eV, corresponding to a frequency of 1420 MHz. In the presence of
the CMB alone, the spin states reach thermal equilibrium with $T_S=T_{\rm
CMB}=2.725 (1+z)$ K on a time-scale of $T_*/(T_{\rm CMB} A_{10}) \simeq 3
\times 10^5 (1+z)^{-1}$ yr, where $A_{10}=2.87 \times 10^{-15}$ s$^{-1}$ is
the spontaneous decay rate of the hyperfine transition. This time-scale is
much shorter than the age of the universe at all redshifts after
cosmological recombination.

The IGM is observable when the kinetic temperature $T_k$ of the gas
differs from $T_{\rm CMB}$ and an effective mechanism couples $T_S$ to
$T_k$. Collisional de-excitation of the triplet level (Purcell \&
Field 1956) dominates at very high redshift, when the gas density (and
thus the collision rate) is still high, but once a significant galaxy
population forms in the universe, the spin temperature is affected
also by an indirect mechanism that acts through the scattering of
Ly$\alpha$ photons (Wouthuysen 1952; Field 1958). Continuum UV photons
produced by early radiation sources redshift by the Hubble expansion
into the local Ly$\alpha$ line at a lower redshift. These photons mix
the spin states via the Wouthuysen-Field process whereby an atom
initially in the $n=1$ state absorbs a Ly$\alpha$ photon, and the
spontaneous decay which returns it from $n=2$ to $n=1$ can result in a
final spin state which is different from the initial one. Since the
neutral IGM is highly opaque to resonant scattering, and the
Ly$\alpha$ photons receive Doppler kicks in each scattering, the shape
of the radiation spectrum near Ly$\alpha$ is determined by $T_k$
(Field 1959b; but see also Hirata 2006; Chuzhoy \& Shapiro 2006), and
the resulting spin temperature (assuming $T_S \gg T_*$) is then a
weighted average of $T_k$ and $T_{\rm CMB}$ (Field 1959a):
\beq T_S=\frac{T_{\rm CMB} T_k (1+x_{\rm tot}) }{T_k + T_{\rm CMB} 
x_{\rm tot}}\ , \eeq where $x_{\rm tot} = x_{\alpha} + x_c$ is the sum
of the radiative and collisional threshold parameters. These
parameters are
\begin{equation}
x_{\alpha} = {{P_{10} T_\star}\over {A_{10} T_{\rm CMB}}}\ ,
\end{equation}
and 
\begin{equation}
x_c = {{4 \kappa_{1-0}(T_k)\, n_H T_\star}\over {3 A_{10} T_{\rm
CMB}}}\ ,\ \end{equation} where $P_{10}$ is the indirect de-excitation
rate of the triplet $n=1$ state via the Wouthuysen-Field process,
related to the total scattering rate $P_{\alpha}$ of Ly$\alpha$
photons by $P_{10}=4 P_{\alpha}/27$ (Field 1958). Also, the atomic
coefficient $\kappa_{1-0}(T_k)$ is tabulated as a function of $T_k$
(Allison \& Dalgarno 1969; Zygelman 2005). Collisions can be
particularly important in collapsed halos (Iliev et al.\ 2003). Note
that we have adopted the modified notation (i.e., in terms of
$x_\alpha$ and $x_c$) of Barkana \& Loeb (2005b). The coupling of the
spin temperature to the gas temperature becomes substantial when
$x_{\rm tot} \gtrsim 1$; in particular, $x_{\alpha} = 1$ defines the
thermalization rate (Madau et al.\ 1997) of $P_{\alpha}$:
\beq P_{\rm th} \equiv \frac{27 A_{10} T_{\rm CMB}}{4 T_*} \simeq 7.6
\times 10^{-12}\, \left(\frac{1+z}{10}\right)\ {\rm s}^{-1}\ . \eeq

A patch of neutral hydrogen at the mean density and with a uniform
$T_S$ produces (after correcting for stimulated emission) an optical
depth at a present-day (observed) wavelength of $21 (1+z)$ cm, 
\beq \tau(z) = 9.0
\times 10^{-3} \left(\frac{T_{\rm CMB}} {T_S} \right) \left (
\frac{\Omega_b h} {0.03} \right) \left(\frac{\Omm}{0.3}\right)^ {-1/2}
\left(\frac{1+z}{10}\right)^{1/2}\ , \eeq assuming a high redshift
$z\gg1$. The observed spectral intensity $I_{\nu}$ relative to the CMB at a
frequency $\nu$ is measured by radio astronomers as an effective
brightness temperature $T_b$ of blackbody emission at this frequency,
defined using the Rayleigh-Jeans limit of the Planck radiation
formula: $I_{\nu} \equiv 2 k_B T_b \nu^2 / c^2 $.

The brightness temperature through the IGM is $T_b=T_{\rm CMB}
e^{-\tau}+T_S (1-e^{-\tau})$, so the observed differential antenna
temperature of this region relative to the CMB is (Madau et al.\ 1997, with
the $\Omm$ dependence added) \beqa T_b&=&(1+z)^{-1} (T_S-T_{\rm CMB})
(1-e^{-\tau}) \nonumber \\ &\simeq& 28\, {\rm mK}\, \left( \frac{\Omega_b
h} {0.033} \right) \left(\frac{\Omm}{0.27}\right)^ {-1/2} \left( \frac{1+z}
{10} \right)^{1/2} \left( \frac{T_S-T_{\rm CMB}} {T_S} \right)\ , \eeqa
where $\tau \ll 1$ is assumed and $T_b$ has been redshifted to redshift
zero. Note that the combination that appears in $T_b$ is
\begin{equation}
{T_S - T_{\rm CMB} \over T_S} = {x_{\rm tot}\over 1+ x_{\rm tot}}
\left(1 - {T_{\rm CMB}\over T_k} \right)\ .
\end{equation}
In overdense regions, the observed $T_b$ is proportional to the
overdensity, and in partially ionized regions $T_b$ is
proportional to the neutral fraction. Also, if $T_S
\gg T_{\rm CMB}$ then the IGM is observed in emission at a level that
is independent of $T_S$. On the other hand, if $T_S \ll T_{\rm CMB}$
then the IGM is observed in absorption at a level that is enhanced by
a factor of $T_{\rm CMB} / T_S$. As a result, a number of cosmic
events are expected to leave observable signatures in the redshifted
21-cm line, as discussed below in further detail.

\section{Galaxy Formation}

\subsection{Growth of linear perturbations}

\label{sec:lin}

As noted in the Introduction, observations of the CMB show that the
universe at cosmic recombination (redshift $z\sim 10^3$) was
remarkably uniform apart from spatial fluctuations in the energy
density and in the gravitational potential of roughly one part in
$10^5$. The primordial inhomogeneities in the density distribution
grew over time and eventually led to the formation of galaxies as well
as galaxy clusters and large-scale structure. In the early stages of
this growth, as long as the density fluctuations on the relevant
scales were much smaller than unity, their evolution can be understood
with a linear perturbation analysis. 

Different physical processes contributed to the perturbation growth (e.g.,
Peebles 1980; Ma \& Bertschinger 1995). In the absence of other influences,
gravitational forces due to density perturbations imprinted by inflation
would have driven parallel perturbation growth in the dark matter, baryons
and photons. However, since the photon sound speed is of order the speed of
light, the radiation pressure produced sound waves on a scale of order the
horizon and suppressed sub-horizon perturbations in the photon density. The
baryonic pressure similarly suppressed perturbations in the gas below the
(much smaller) so-called baryonic {\it Jeans} scale. Since the formation of
hydrogen at recombination had decoupled the cosmic gas from its mechanical
drag on the CMB, the baryons subsequently began to fall into the
pre-existing gravitational potential wells of the dark matter.

Spatial fluctuations developed in the gas temperature as well as in
the gas density. Both the baryons and the dark matter were affected on
small scales by the temperature fluctuations through the gas pressure.
Compton heating due to scattering of the residual free electrons
(constituting a fraction $\sim10^{-4}$) with the CMB photons remained
effective, keeping the gas temperature fluctuations tied to the photon
temperature fluctuations, even for a time after recombination. The
growth of linear perturbations can be calculated with the standard
CMBFAST code (Seljak \& Zaldarriaga 1996; see http://www.cmbfast.org),
after a modification to account for the fact that the speed of sound
of the gas also fluctuates spatially (Yamamoto \etal 1997, 1998; Naoz
\& Barkana 2005). 

\begin{figure}
\includegraphics[width=5in]{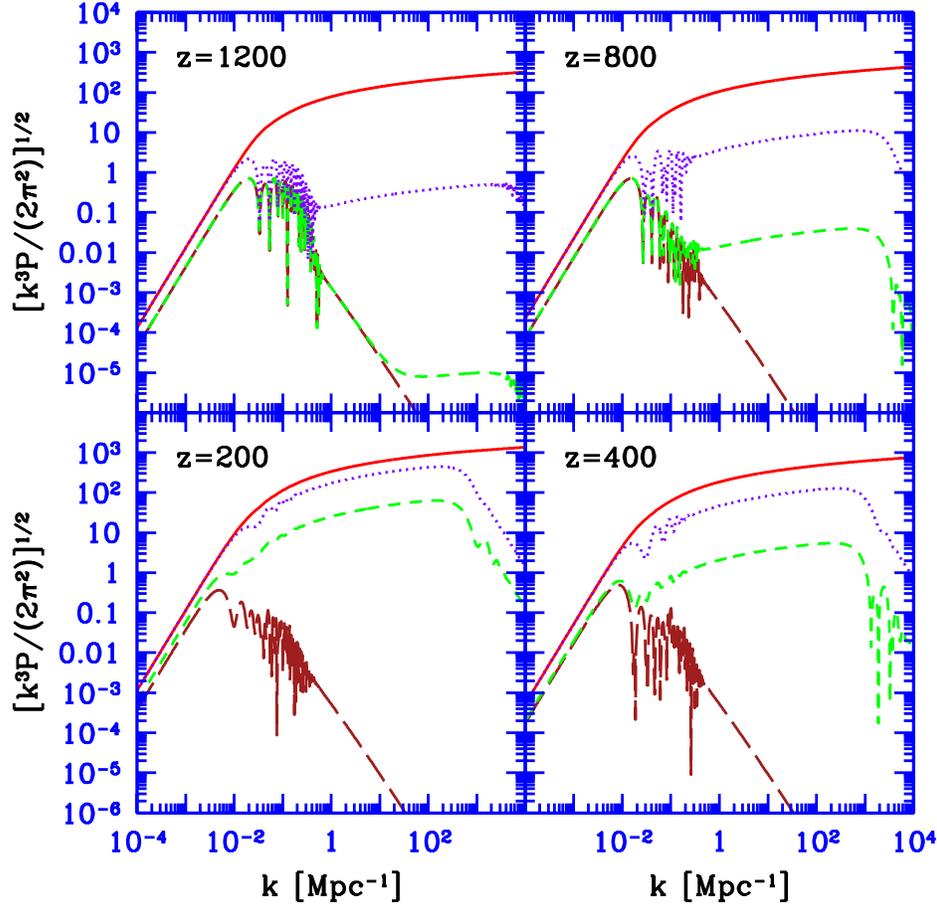}
\caption{Power spectra of density and temperature fluctuations vs.\
comoving wavenumber, at redshifts 1200, 800, 400, and 200, adopted
from Naoz \& Barkana 2005. We consider fluctuations in the CDM density
(solid curves), baryon density (dotted curves), baryon temperature
(short-dashed curves), and photon temperature (long-dashed curves).}
\label{fig:photons}
\end{figure}

The magnitude of the fluctuations in the CDM and baryon densities, and
in the baryon and photon temperatures, is shown in
Figure~\ref{fig:photons}, in terms of the dimensionless combination
$[k^3 P(k)/(2 \pi^2)]^{1/2}$, where $P(k)$ is the corresponding power
spectrum of fluctuations in terms of the comoving wavenumber $k$ of
each Fourier mode. After recombination, two main drivers affect the
baryon density and temperature fluctuations, namely, the
thermalization with the CMB and the gravitational force that attracts
the baryons to the dark matter potential wells. As shown in the
figure, the density perturbations in all species grow together on
scales where gravity is unopposed, outside the horizon (i.e., at $k
\lesssim 0.01$ Mpc$^{-1}$ at $z \sim 1000$). At $z=1200$ the
perturbations in the baryon-photon fluid oscillate as acoustic waves
on scales of order the sound horizon ($k \sim 0.01~{\rm Mpc^{-1}}$),
while smaller-scale perturbations in both the photons and baryons are
damped by photon diffusion (Silk damping) and the drag of the
diffusing photons on the baryons. On sufficiently small scales the
power spectra of baryon density and temperature roughly assume the
shape of the dark matter fluctuations (except for the gas-pressure
cutoff at the very smallest scales), due to the effect of
gravitational attraction on the baryon density and of the resulting
adiabatic expansion on the gas temperature. After the mechanical
coupling of the baryons to the photons ends at $z \sim 1000$, the
baryon density perturbations gradually grow towards the dark matter
perturbations because of gravity. Similarly, after the thermal
coupling ends at $z \sim 200$, the baryon temperature fluctuations are
driven by adiabatic expansion towards a value of 2/3 of the baryon
density fluctuations. As the figure shows, by $z=200$ the baryon
infall into the dark matter potentials is well advanced and adiabatic
expansion is becoming increasingly important in setting the baryon
temperature.

\subsection{Formation of the First Stars}

Theoretical expectations for the properties of the first galaxies
(Barkana \& Loeb 2001) are based on the standard cosmological model
outlined in the Introduction. The formation of the first bound objects
marked the central milestone in the transition from the initial
simplicity (discussed in the previous subsection) to the present-day
complexity. Stars and quasars output copious radiation and also
produced explosions and outflows that brought into the IGM chemical
products from stellar nucleosynthesis and enhanced magnetic
fields. However, the formation of the very first stars, in a universe
that had not yet suffered such feedback, remains a well-specified
problem for theorists.

Stars form when huge amounts of matter collapse to enormous
densities. However, the process can be stopped if the pressure exerted by
the hot intergalactic gas prevents outlying gas from falling into dark
matter concentrations. As the gas falls into a dark matter halo, it forms
shocks due to converging supersonic flows and in the process heats up and
can only collapse further by first radiating its energy away. This
restricts this process of collapse to very large clumps of dark matter that
are around $100\,000$ times the mass of the Sun. Inside these clumps, the
shocked gas loses energy by emitting radiation from excited molecular
hydrogen that formed naturally within the primordial gas mixture of
hydrogen and helium (Peebles 1984; Haiman \etal 1996; Tegmark \etal 1997).

The first stars are expected to have been quite different from the
stars that form today in the Milky Way. The higher pressure within the
primordial gas due to the presence of fewer cooling agents suggests
that fragmentation only occurred into relatively large units, in which
gravity could overcome the pressure. Due to the lack of carbon,
nitrogen, and oxygen -- elements that would normally dominate the
nuclear energy production in modern massive stars -- the first stars
must have condensed to extremely high densities and temperatures
before nuclear reactions were able to heat the gas and balance
gravity. These unusually massive stars produced high luminosities of
UV photons, but their nuclear fuel was exhausted after 2--3 million
years, resulting in a huge supernova or in collapse to a massive black
hole. For additional details about the properties of the first stars,
see the review by Bromm \& Larson (2004).

Advances in computing power have made possible detailed numerical
simulations of how the first stars formed (Bromm \etal 2002; Abel \etal
2002; Bromm \& Loeb 2004; Yoshida \etal 2006). These simulations begin in
the early universe, in which dark matter and gas are distributed uniformly,
apart from tiny variations in density and temperature that are
statistically distributed according to the patterns observed in the CMB. In
order to span the vast range of scales needed to simulate an individual
star within a cosmological context, the latest code (Yoshida \etal 2006)
follows a box 0.3 Mpc in length and zooms in repeatedly on the densest part
of the first collapsing cloud that is found within the simulated
volume. The simulation follows gravity, hydrodynamics, and chemical
processes in the primordial gas, and resolves a scale 10 orders of
magnitudes smaller than that of the simulated box. While the resolved scale
is still three orders of magnitudes larger than the size of the Sun, these
simulations have established that the first stars formed within halos
containing $\sim 10^5 M_{\odot}$ in total mass, and indicate that the first
stars most likely weighed $\sim 100 M_{\odot}$ each.

\begin{figure}
\centering
\includegraphics[width=5in]{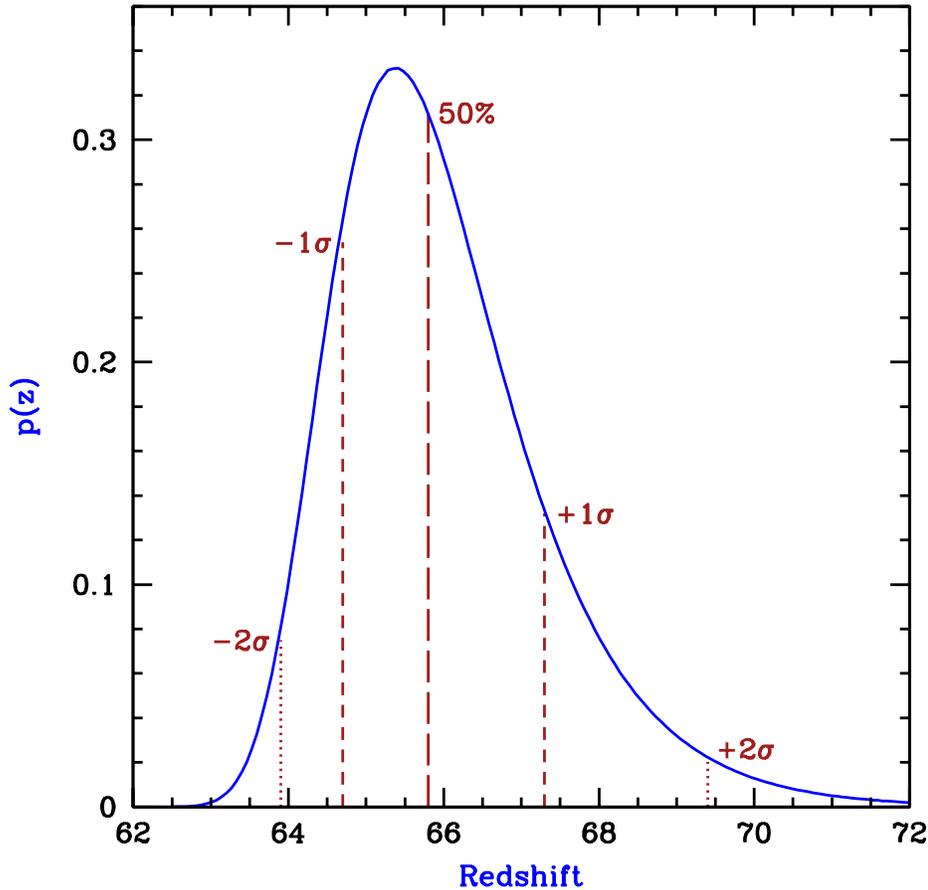}
\caption{The probability of finding the first observable star in the
Universe, as a function of redshift (from Naoz \etal 2006). Vertical lines
show the median redshift as well as the central $1-\sigma$ $(68\%)$ range
and $2-\sigma$ ($95\%$) range. This result is based on an analytical model
that accounts for the size of the universe and for three essential
ingredients: the light travel time from distant galaxies, Poisson and
density fluctuations on all scales, and the effect of very early cosmic
history on galaxy formation.}
\label{fig:1stStar}
\end{figure}

To estimate {\it when}\/ the first stars formed we must remember that the
first $100\,000$ solar mass halos collapsed in regions that happened to
have a particularly high density enhancement very early on. There was
initially only a small abundance of such regions in the entire universe, so
a simulation that is limited to a small volume is unlikely to find such
halos until much later. Simulating the entire universe is well beyond the
capabilities of current simulations, but analytical models predict that the
first observable star in the universe (Figure~\ref{fig:1stStar}) probably
formed 30 million years after the Big Bang (Naoz \etal 2006), less than a
quarter of one percent of the Universe's total age of 13.7 billion years.

Although stars were extremely rare at first, gravitational collapse
increased the abundance of galactic halos and star formation sites
with time (Figure~\ref{fig:history}). Radiation from the first stars
is expected to have eventually dissociated all the molecular hydrogen
in the intergalactic medium, leading to the domination of a second
generation of larger galaxies where the gas cooled via radiative
transitions in atomic hydrogen and helium (Haiman \etal 1997). Atomic
cooling occurred in halos of mass above $\sim10^8 M_{\odot}$, in which
the infalling gas was heated above 10,000 K and became ionized. The
first galaxies to form through atomic cooling are expected to have
formed around redshift 45 (Naoz \etal 2006), and such galaxies were
likely the main sites of star formation by the time reionization began
in earnest. As the IGM was heated above 10,000 K by reionization, its
pressure jumped and prevented the gas from accreting into newly
forming halos below $\sim10^9 M_{\odot}$ (Rees 1986). The first
Milky-Way-sized halo $M = 10^{12} M_{\odot}$ is predicted to have
formed 400 million years after the Big Bang (Naoz \etal 2006), but
such halos have become typical galactic hosts only in the last five
billion years.

\subsection{Gamma-ray Bursts: Probing the First Stars One 
Star at a Time}

\begin{figure}
\centering
\includegraphics[height=6cm]{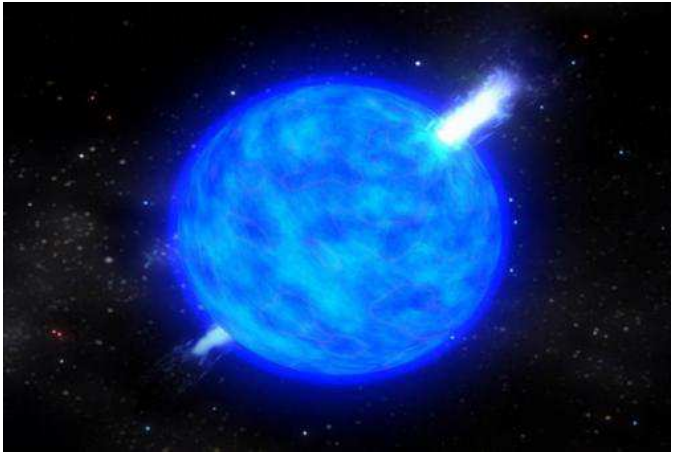}
\caption{Illustration of a long-duration gamma-ray burst in the popular
``collapsar'' model (Zhang \etal 2003).
The collapse of the core of a massive star (which lost its hydrogen
envelope) to a black hole generates two opposite jets moving out at a
speed close to the speed of light. The jets drill a hole in the star
and shine brightly towards an observer who happens to be located
within with the collimation cones of the jets. The jets emanating from
a single massive star are so bright that they can be seen across the
Universe out to the epoch when the first stars formed. Upcoming
observations by the {\it Swift}\/ satellite will have the sensitivity
to reveal whether Pop~III stars served as progenitors of gamma-ray
bursts (for more information see http://swift.gsfc.nasa.gov/).}
\label{grb}
\end{figure}

Gamma-Ray Bursts (GRBs) are believed to originate in compact remnants
(neutron stars or black holes) of massive stars. Their high
luminosities make them detectable out to the edge of the visible
Universe (Lamb \& Reichart 2000; Ciardi \& Loeb 2000). GRBs offer the
opportunity to detect the most distant (and hence earliest) population
of massive stars, the so-called Population~III (or Pop~III), one star
at a time (Figure~\ref{grb}). In the hierarchical assembly process of
halos that are dominated by cold dark matter (CDM), the first galaxies
should have had lower masses (and lower stellar luminosities) than
their more recent counterparts. Consequently, the characteristic
luminosity of galaxies or quasars is expected to decline with
increasing redshift (but note that star formation may favor more
massive halos due to feedback effects, similar to the "downsizing"
observed at low redshift; e.g., Bundy et al.\ 2006). GRB afterglows,
which already produce a peak flux comparable to that of quasars or
starburst galaxies at $z\sim 1-2$, are therefore expected to outshine
any competing source at the highest redshifts, when the first dwarf
galaxies formed in the Universe.

GRBs, the electromagnetically-brightest explosions in the Universe,
should be detectable out to redshifts $z>10$. High-redshift GRBs can
be identified through infrared photometry, based on the Ly$\alpha$
break induced by absorption of their spectrum at wavelengths below
$1.216\, \mu {\rm m}\, [(1+z)/10]$. Follow-up spectroscopy of
high-redshift candidates can then be performed on a 10-meter-class
telescope. GRB afterglows offer the opportunity to detect stars as
well as to probe the metal enrichment level (Furlanetto \& Loeb 2003)
of the intervening IGM. Recently, the ongoing {\it Swift} mission
(Gehrels \etal 2004) has detected a GRB originating at $z\simeq 6.3$
(e.g., Haislip \etal 2006), thus demonstrating the viability of GRBs
as probes of the early Universe.

\begin{figure}
\centering
\includegraphics[width=5in]{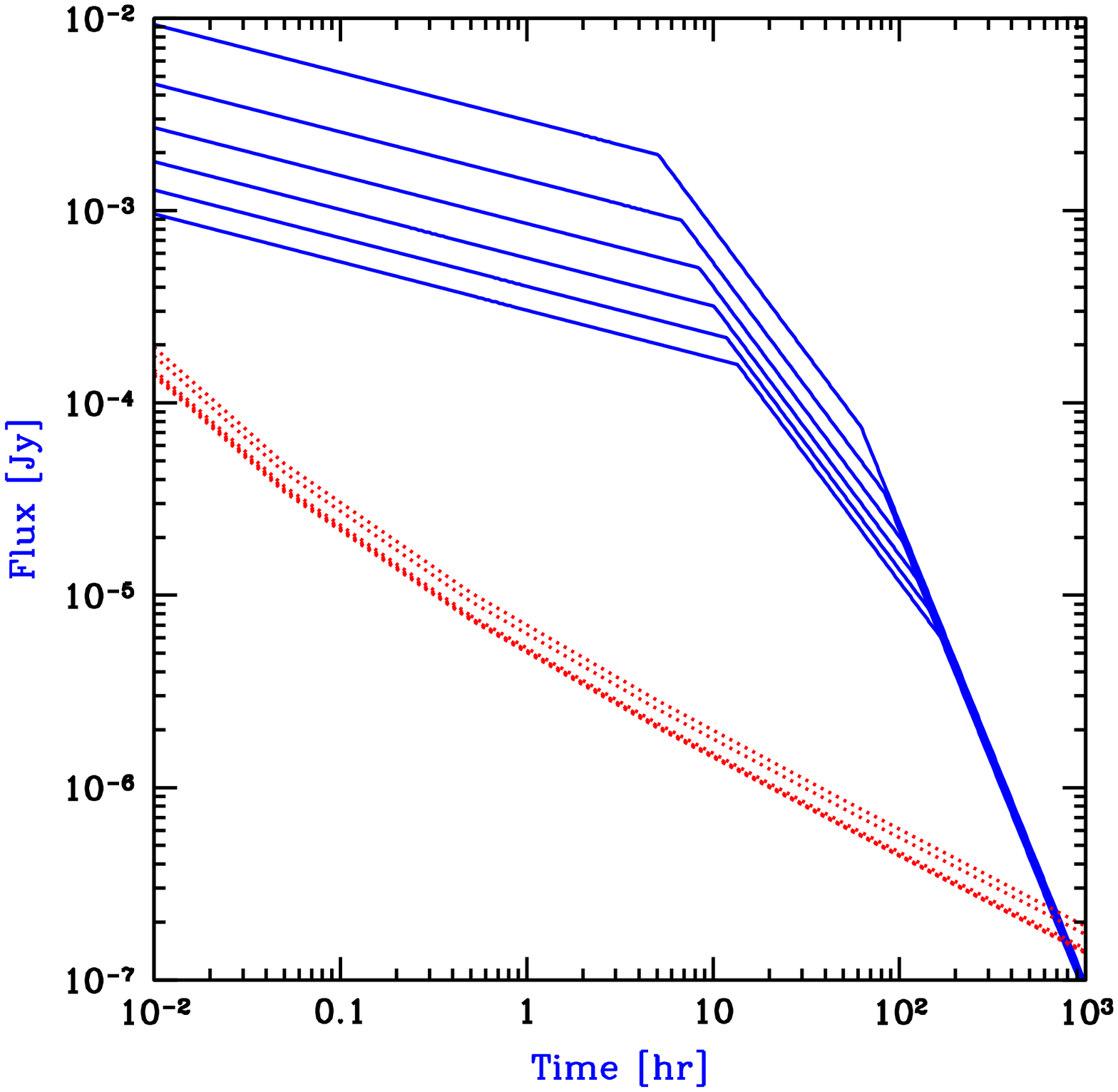}
\caption{GRB afterglow flux as a function of time since the $\gamma$-ray
trigger in the observer frame (taken from Barkana \& Loeb 2004a). The
flux (solid curves) is calculated at the redshifted Ly$\alpha$
wavelength. The dotted curves show the planned detection threshold for
the {\it James Webb Space Telescope} ({\it JWST}), assuming a spectral
resolution $R=5000$ with the near infrared spectrometer, a signal to
noise ratio of 5 per spectral resolution element, and an exposure time
equal to $20\%$ of the time since the GRB explosion (see
http://www.ngst.stsci.edu/nms/main/~).  Each set of curves shows a
sequence of redshifts, namely $z=5$, 7, 9, 11, 13, and 15,
respectively, from top to bottom.}
\label{fig:GRB}
\end{figure}

Another advantage of GRBs is that the GRB afterglow flux at a given
observed time lag after the $\gamma$-ray trigger is not expected to fade
significantly with increasing redshift, since higher redshifts translate to
earlier times in the source frame, during which the afterglow is
intrinsically brighter (Ciardi \& Loeb 2000). For standard afterglow
lightcurves and spectra, the increase in the luminosity distance with
redshift is compensated by this {\it cosmological time-stretching} effect
(Barkana \& Loeb 2004a) as shown in Figure~\ref{fig:GRB}.

GRB afterglows have smooth (broken power-law) continuum spectra unlike
quasars which show strong spectral features (such as broad emission lines
or the so-called ``blue bump'') that complicate the extraction of IGM
absorption features. In particular, the continuum extrapolation into the
Ly$\alpha$ damping wing (see section~\ref{sec:absorb}) during the epoch of
reionization is much more straightforward for the smooth UV spectra of GRB
afterglows than for quasars with an underlying broad Ly$\alpha$ emission
line (Barkana \& Loeb 2004a). However, the interpretation may be
complicated by the presence of damped Ly$\alpha$ absorption by dense
neutral hydrogen in the immediate environment of the GRB within its host
galaxy. Since GRBs originate from the dense environment of active star
formation, such damped absorption is expected and indeed has been seen,
including in the most distant GRB at $z=6.3$ (Totani \etal 2006).

\subsection{The epoch of reionization}

\begin{figure}
\centering
\includegraphics[width=2.5in]{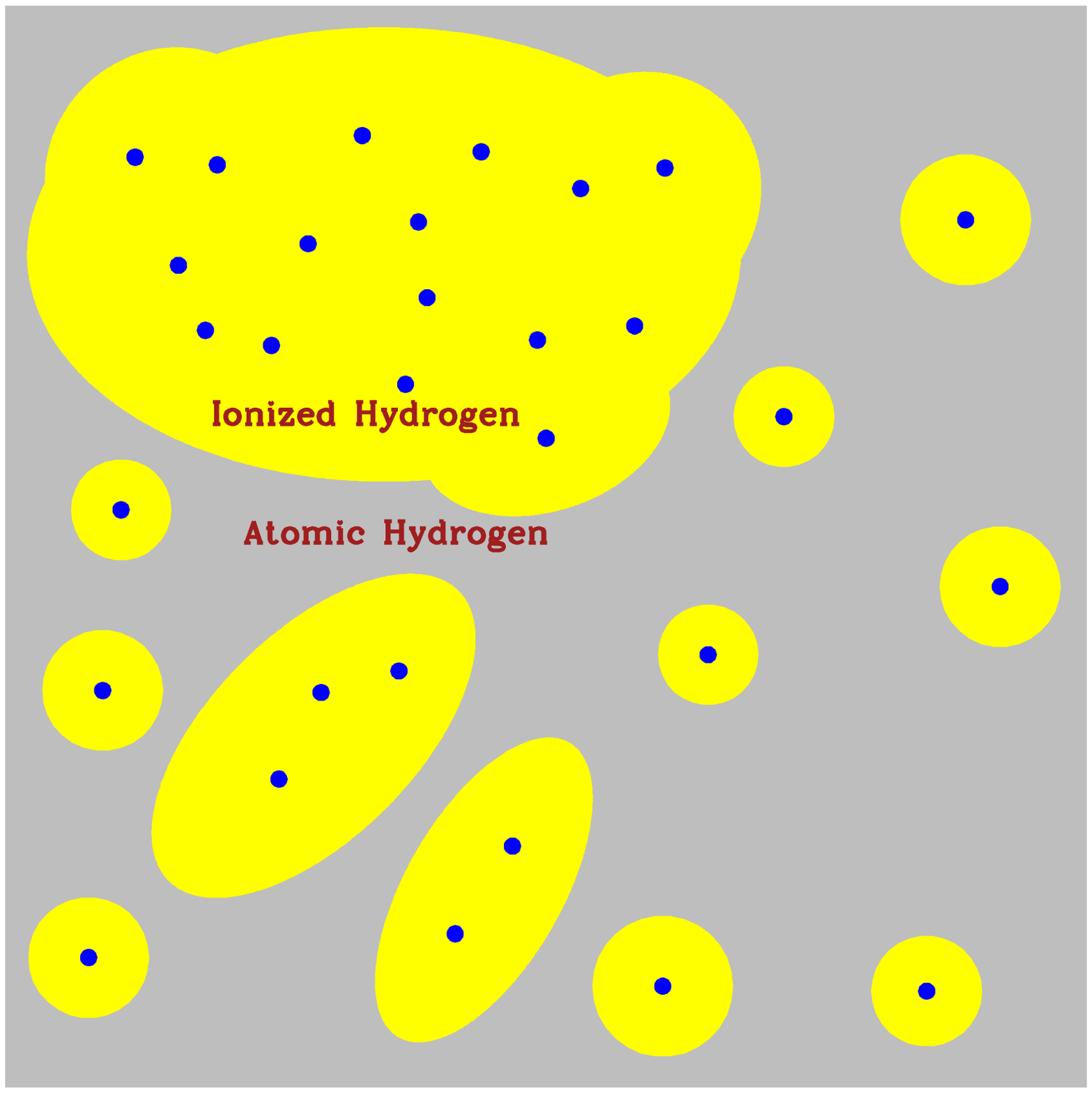}
\hfill
\includegraphics[width=2.5in]{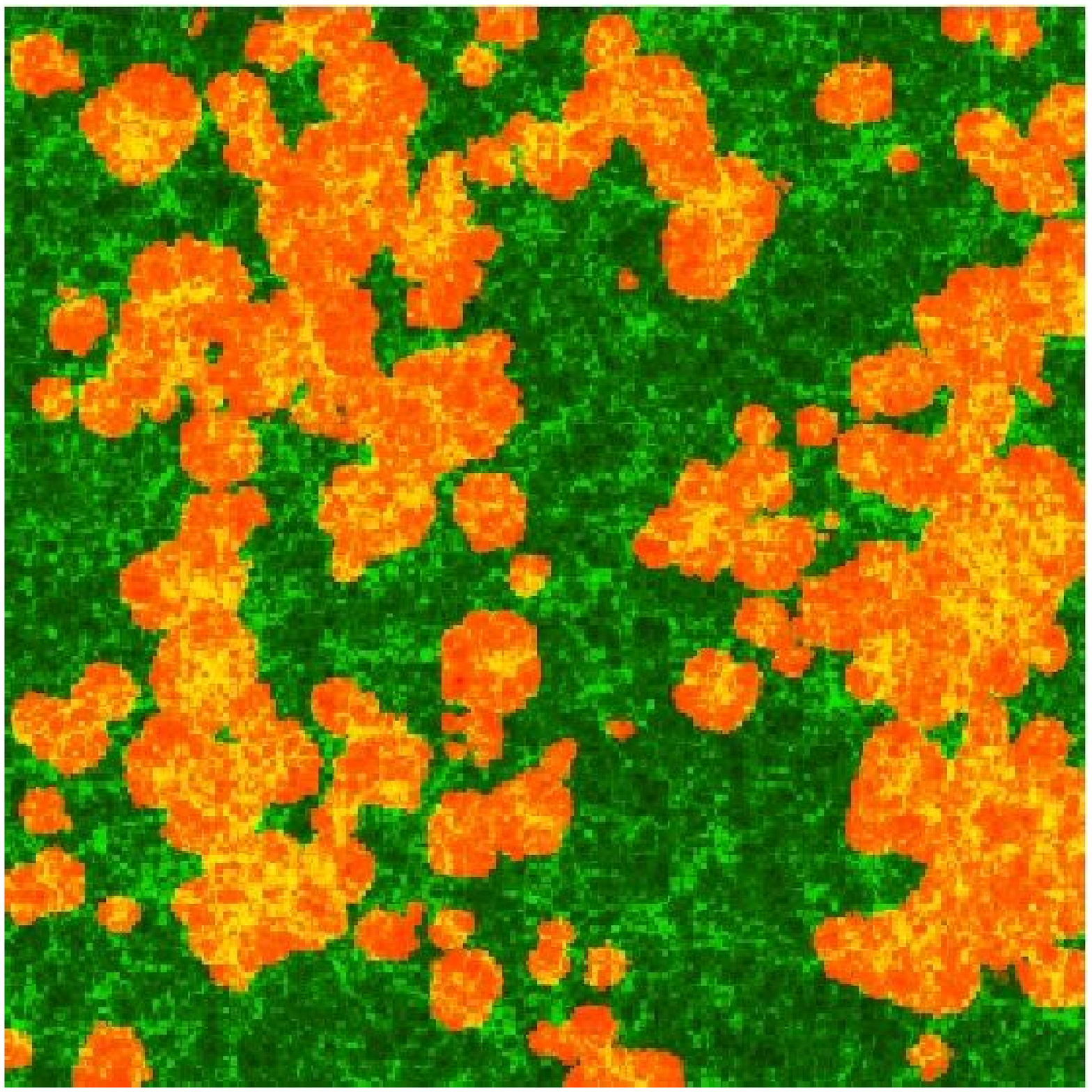}
\caption{The spatial structure of cosmic reionization. The illustration 
(left panel, based on Barkana \& Loeb 2004b) shows how regions with
large-scale overdensities form large concentrations of galaxies (dots)
whose ionizing photons produce enormous joint ionized bubbles (upper
left). At the same time, galaxies are rare within large-scale voids,
in which the IGM is still mostly neutral (lower right). A numerical
simulation of reionization (right panel, from Mellema \etal 2006)
indeed displays such variation in the sizes of ionized bubbles
(orange), shown overlayed on the density distribution (green).}
\label{fig:rei}
\end{figure}

Given the understanding described above of how many galaxies formed at
various times, the course of reionization can be determined
universe-wide by counting photons from all sources of light (Arons \&
Wingert 1972; Shapiro \& Giroux 1987; Meiksin \& Madau 1993;
Miralda-Escud\'e \& Ostriker 1990; Tegmark \etal 1994; Kamionkowski
\etal 1994; Fukugita \& Kawasaki 1994; Shapiro \etal 1994; Haiman \&
Loeb 1997). Both stars and black holes contribute ionizing photons,
but the early universe is dominated by small galaxies which in the
local universe have central black holes that are disproportionately
small, and indeed quasars are rare above redshift 6 (Fan \etal
2003). Thus, stars most likely dominated the production of ionizing UV
photons during the reionization epoch [although high-redshift galaxies
should have also emitted X-rays from accreting black holes and
accelerated particles in collisionless shocks (Oh 2001)]. Since most
stellar ionizing photons are only slightly more energetic than the
13.6 eV ionization threshold of hydrogen, they are absorbed
efficiently once they reach a region with substantial neutral hydrogen
(e.g., see section~(\ref{sec:absorb})). This makes the IGM during
reionization a two-phase medium characterized by highly ionized
regions separated from neutral regions by sharp ionization fronts (see
Figure~\ref{fig:rei}).

We can obtain a first estimate of the requirements of reionization by
demanding one stellar ionizing photon for each hydrogen atom in the
IGM. If we conservatively assume that stars within the early galaxies
were similar to those observed locally, then each star produced $\sim
4000$ ionizing photons per baryon. Star formation is observed today to
be an inefficient process, but even if stars in galaxies formed out of
only $\sim10\%$ of the available gas, it was still sufficient to
accumulate a small fraction (of order $0.1\%$) of the total baryonic
mass in the universe into galaxies in order to reionize the entire
IGM. More accurate estimates of the actual required fraction account
for the formation of some primordial stars (which were massive,
efficient ionizers, as discussed above), and for recombinations of
hydrogen atoms at high redshifts and in dense regions.

From studies of quasar absorption lines at $z\sim 6$ we know that the
IGM is highly ionized a billion years after the big bang (see the
review by Fan \etal 2006b). There are hints, however, that some large
neutral hydrogen regions persist at these early times (Wyithe \& Loeb
2004a; Mesinger \& Haiman 2004; Lidz \etal 2006) and so this suggests
that we may not need to go to much higher redshifts to begin to see
the epoch of reionization.  (Interestingly, these inferences were
deduced long before the latest WMAP results were announced, when it
was widely believed that reionization occurred much earlier.)  We now
know that the universe could not have fully reionized earlier than an
age of 300 million years, since WMAP observed the effect of the
freshly created plasma at reionization on the large-scale polarization
anisotropies of the CMB and this limits the reionization redshift
(Spergel \etal 2006); an earlier reionization, when the universe was
denser, would have created a stronger scattering signature that would
be inconsistent with the WMAP observations. In any case, the redshift
at which reionization ended only constrains the overall cosmic
efficiency of ionizing photon production. In comparison, a detailed
picture of reionization as it happens will teach us a great deal about
the population of young galaxies that produced this cosmic phase
transition.

A key point is that the spatial distribution of ionized bubbles is
determined by clustered groups of galaxies and not by individual
galaxies. At such early times galaxies were strongly clustered even on
very large scales (up to tens of Mpc), and these scales therefore
dominate the structure of reionization (Barkana \& Loeb 2004b). The
basic idea is simple (Kaiser 1984). At high redshift, galactic halos
are rare and correspond to rare, high density peaks. As an analogy,
imagine searching on Earth for mountain peaks above 5000 meters. The
200 such peaks are not at all distributed uniformly but instead are
found in a few distinct clusters on top of large mountain
ranges. Given the large-scale boost provided by a mountain range, a
small-scale crest need only provide a small additional rise in order
to become a 5000 meter peak. The same crest, if it formed within a
valley, would not come anywhere near 5000 meters in total
height. Similarly, in order to find the early galaxies, one must first
locate a region with a large-scale density enhancement, and then
galaxies will be found there in abundance.

The ionizing radiation emitted from the stars in each galaxy initially
produces an isolated ionized bubble. However, in a region dense with
galaxies the bubbles quickly overlap into one large bubble, completing
reionization in this region while the rest of the universe is still
mostly neutral (Figure~\ref{fig:rei}). Most importantly, since the
abundance of rare density peaks is very sensitive to small changes in
the density threshold, even a large-scale region with a small enhanced
density (say, 10\% above the mean density of the universe) can have a
much larger concentration of galaxies than in other regions (e.g., a
50\% enhancement). On the other hand, reionization is harder to
achieve in dense regions, since the protons and electrons collide and
recombine more often in such regions, and newly-formed hydrogen atoms
need to be reionized again by additional ionizing photons. However,
the overdense regions end up reionizing first since the number of
ionizing sources in these regions is increased so strongly (Barkana \&
Loeb 2004b). The large-scale topology of reionization is therefore
inside out, with underdense voids reionizing only at the very end of
reionization, with the help of extra ionizing photons coming in from
their surroundings (which have a higher density of galaxies than the
voids themselves). This is a key prediction awaiting observational
testing.

\begin{figure}
\centering
\includegraphics[width=5.3in]{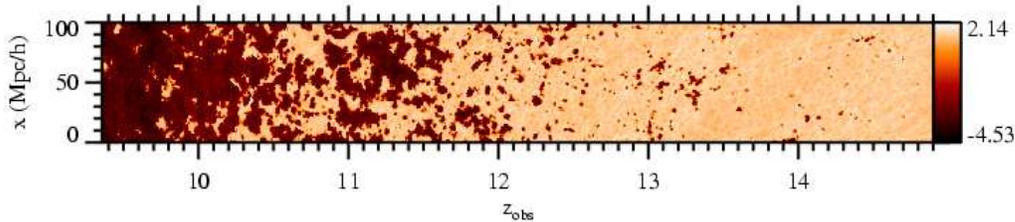}
\caption{Close-up of cosmic evolution during the epoch of reionization, as
revealed in a predicted 21-cm map of the IGM based on a numerical
simulation (from Mellema \etal 2006). This map is constructed from slices
of the simulated cubic box of side 150 Mpc (in comoving units), taken at
various times during reionization, which for the parameters of this
particular simulation spans a period of 250 million years from redshift 15
down to 9.3. The vertical axis shows position $\chi$ in units of Mpc/h
(where $h=0.7$). This two-dimensional slice of the sky (one linear
direction on the sky versus the line-of-sight or redshift direction) shows
$\log_{10}(T_b)$, where $T_b$ (in mK) is the 21-cm brightness temperature
relative to the CMB. Since neutral regions correspond to strong emission
(i.e., a high $T_b$), this slice illustrates the global progress of
reionization and the substantial large-scale spatial fluctuations in
reionization history. Observationally it corresponds to a narrow strip half
a degree in length on the sky observed with radio telescopes over a
wavelength range of 2.2 to 3.4 m (with each wavelength corresponding to
21-cm emission at a specific line-of-sight distance and redshift).}
\label{fig:Mellema}
\end{figure}

Detailed analytical models that account for large-scale variations in
the abundance of galaxies (Furlanetto \etal 2004) confirm that the
typical bubble size starts well below a Mpc early in reionization, as
expected for an individual galaxy, rises to 5--10 Mpc during the
central phase (i.e., when the universe is half ionized), and then by
another factor of $\sim$5 towards the end of reionization. These
scales are given in comoving units that scale with the expansion of
the universe, so that the actual sizes at a redshift $z$ were smaller
than these numbers by a factor of $1+z$. Numerical simulations have
only recently begun to reach the enormous scales needed to capture
this evolution (Ciardi \etal 2003; Mellema \etal 2006; Zahn
\etal 2006). Accounting precisely for gravitational evolution on a
wide range of scales but still crudely for gas dynamics, star
formation, and the radiative transfer of ionizing photons, the
simulations confirm that the large-scale topology of reionization is
inside out, and that this topology can be used to study the abundance
and clustering of the ionizing sources (Figures~\ref{fig:rei} and
\ref{fig:Mellema}).

Wyithe \& Loeb (2004b) showed that the characteristic size of the ionized
bubbles at the end reionization can be calculated based on simple
considerations that only depend on the power-spectrum of density
fluctuations and the redshift. As the size of an ionized bubble increases,
the time it takes a 21-cm photon to traverse it gets longer. At the same
time, the variation in the time at which different regions reionize becomes
smaller as the regions grow larger. Thus, there is a maximum size above
which the photon crossing time is longer than the cosmic variance in
ionization time. Regions bigger than this size will be ionized at their
near side by the time a 21-cm photon will cross them towards the observer
from their far side. They would appear to the observer as one-sided, and
hence signal the end of reionization. These ``light cone'' considerations
imply a characteristic size for the ionized bubbles of $\sim 10$ physical
Mpc at $z\sim 6$ (equivalent to 70 Mpc today).  This result implies that
future radio experiments should be tuned to a characteristic angular scale
of tens of arcminutes and have a minimum frequency band-width of 5-10 MHz
for an optimal detection of 21-cm brightness fluctuations near the end of
reionization.

\subsection{Post-reionization suppression of Low Mass Galaxies}

After the ionized bubbles overlapped in each region, the ionizing
background increased sharply, and the IGM was heated by the ionizing
radiation to a temperature $T_{\rm IGM}\gtrsim 10^4$ K. Due to the
substantial increase in the IGM pressure, the smallest mass scale into
which the cosmic gas could fragment, the so-called Jeans mass,
increased dramatically, changing the minimum mass of forming galaxies
(Rees 1986; Efstathiou 1992; Gnedin \& Ostriker 1997; Miralda-Escud\'e
\& Rees 1998; Benson et al.\ 2003).

Gas infall depends sensitively on the Jeans mass. When a halo more massive
than the Jeans mass begins to form, the gravity of its dark matter
overcomes the gas pressure. Even in halos below the Jeans mass, although
the gas is initially held up by pressure, once the dark matter collapses
its increased gravity pulls in some gas (Haiman \etal 1996). Thus, the
Jeans mass is generally higher than the actual limiting mass for
accretion. Before reionization, the IGM is cold and neutral, and the Jeans
mass plays a secondary role in limiting galaxy formation compared to
cooling. After reionization, the Jeans mass is increased by several orders
of magnitude due to the photoionization heating of the IGM, and hence
begins to play a dominant role in limiting the formation of stars.  Gas
infall in a reionized and heated Universe has been investigated in a number
of numerical simulations. Thoul \& Weinberg (1996) inferred, based on a
spherically-symmetric collapse simulation, a reduction of $\sim 50\%$ in
the collapsed gas mass due to heating, for a halo of circular velocity
$V_c\sim 50\ {\rm km\ s}^{-1}$ at $z=2$, and a complete suppression of
infall below $V_c \sim 30\ {\rm km\ s}^{-1}$. Kitayama \& Ikeuchi (2000)
also performed spherically-symmetric simulations but included
self-shielding of the gas, and found that it lowers the circular velocity
thresholds by $\sim 5\ {\rm km\ s}^{-1}$.  Three dimensional numerical
simulations (Quinn \etal 1996; Weinberg \etal 1997; Navarro \& Steinmetz
1997) found a significant suppression of gas infall in even larger halos
($V_c \sim 75\ {\rm km\ s}^{-1}$), but this was mostly due to a suppression
of late infall at $z\la 2$.

When a volume of the IGM is ionized by stars, the gas is heated to a
temperature $T_{\rm IGM}\sim 10^4$ K. If quasars dominate the UV background
at reionization, their harder photon spectrum leads to $T_{\rm IGM}>
2\times 10^4$ K. Including the effects of dark matter, a given temperature
results in a linear Jeans mass corresponding to a halo circular velocity of
\beq V_J\approx 80 \left(\frac{T_{\rm IGM}}{1.5\times 10^4 {\rm
K}}\right)^{1/2}\ {\rm km\ s}^{-1}. \eeq In halos with a circular
velocity well above $V_J$, the gas fraction in infalling gas equals
the universal mean of $\Omega_b/\Omega_m$, but gas infall is
suppressed in smaller halos.  A simple estimate of the limiting
circular velocity, below which halos have essentially no gas infall,
is obtained by substituting the virial overdensity for the mean
density in the definition of the Jeans mass. The resulting estimate is
\beq V_{\rm lim}=34 \left(\frac{T_{\rm IGM}}{1.5\times 10^4 {\rm
K}}\right)^{1/2}\ {\rm km\ s}^{-1}. \eeq This value is in rough
agreement with the numerical simulations mentioned before.  A more
recent study by Dijkstra et al. (2004) indicated that at the high
redshifts of $z>10$ gas could nevertheless assemble into halos with
circular velocities as low as $v_c\sim 10~{\rm km~s^{-1}}$, even
sometime after the establishment of a UV background.

Although the Jeans mass is closely related to the rate of gas infall at a
given time, it does not directly yield the total gas residing in halos at a
given time. The latter quantity depends on the entire history of gas
accretion onto halos, as well as on the merger histories of halos, and an
accurate description must involve a time-averaged Jeans mass. 
The gas content of halos in simulations is well fit by an expression which
depends on the filtering mass, a particular time-averaged Jeans mass
(Gnedin \& Hui 1998; Gnedin 2000).

The reionization process was not perfectly synchronized throughout the
Universe. Large-scale regions with a higher density than the mean
tended to form galaxies first and reionized earlier than underdense
regions. The suppression of low-mass galaxies by reionization is
therefore modulated by the fluctuations in the timing of reionization
(Babich \& Loeb 2005).  Inhomogeneous reionization imprints a
signature on the power-spectrum of low-mass galaxies. Future
high-redshift galaxy surveys hoping to constrain inflationary
parameters must properly model the effects of reionization;
conversely, they will also place new constraints on the thermal
history of the IGM during reionization.

\section{21-cm cosmology}

\subsection{A handy tool for studying cosmic reionization}

The prospect of studying reionization by mapping the distribution of atomic
hydrogen across the universe using its prominent 21-cm spectral line has
motivated several teams to design and construct arrays of low-frequency
radio telescopes; the Low Frequency Array (http://www.lofar.org/), the
Mileura Wide-Field Array ({\it
http://www.haystack.mit.edu/ast/arrays/mwa/site/index.html}), the Primeval
Structure Telescope ({\it http://arxiv.org/abs/astro-ph/0502029}), and
ultimately the Square Kilometer Array ({\it http://www.skatelescope.org})
will search over the next decade for 21-cm emission or absorption from
$z\sim 6.5$--15, redshifted and observed today at relatively low
frequencies which correspond to wavelengths of 1.5 to 4 meters.

\begin{figure}
\centering
\includegraphics[width=4.8in]{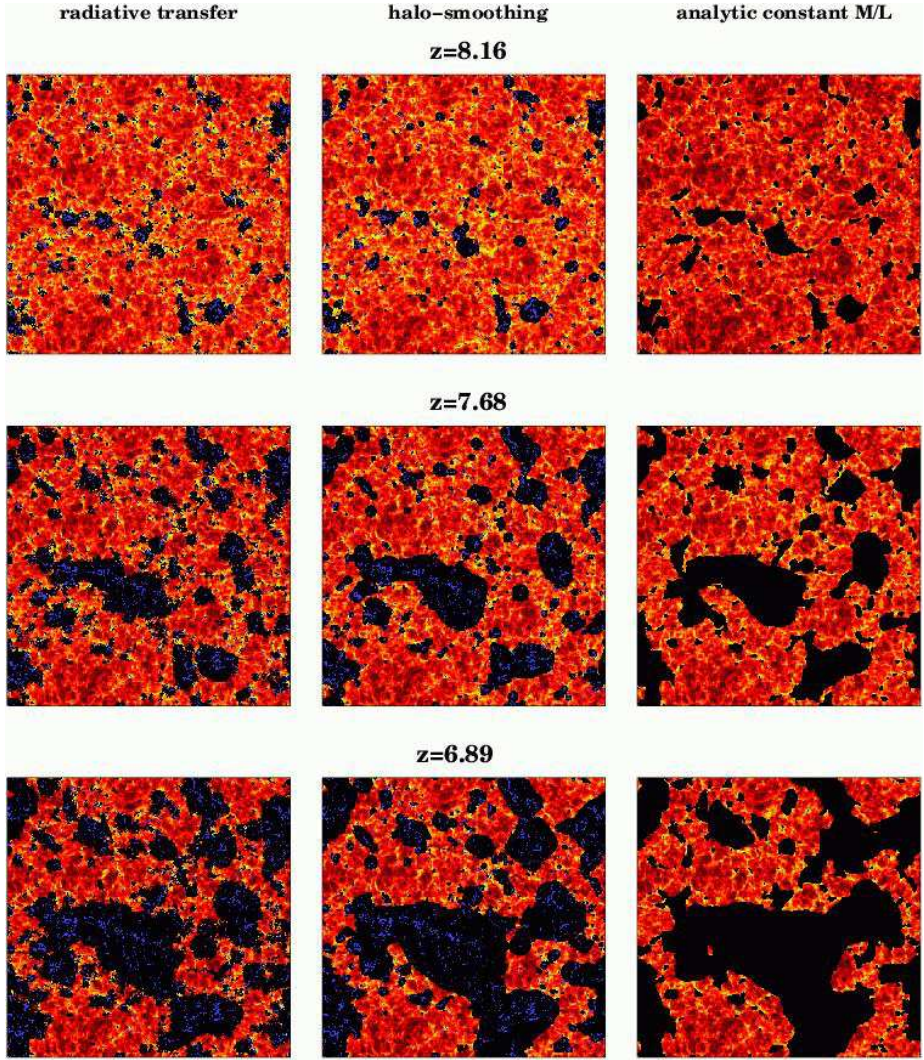}
\caption{Maps of the 21-cm brightness temperature comparing results of a
numerical simulation and of two simpler numerical schemes, at three
different redshifts (from Zahn \etal 2006). Each map is 65.6 Mpc/$h$ on a
side, with a depth (0.25 Mpc/$h$) that is comparable to the frequency
resolution of planned experiments. The ionized fractions are $x_{\rm
i}=0.13$, 0.35 and 0.55 for $z=8.16$, 7.26 and 6.89, respectively. All
three maps show a very similar large-scale ionization topology. \emph{Left
column:} Numerical simulation, showing the ionized bubbles (black) produced
by the ionizing sources (blue dots) that form in the
simulation. \emph{Middle column:} Numerical scheme that applies the
Furlanetto \etal (2004) analytical model to the final distribution of
ionizing sources that form in the simulation. \emph{Right column:}
Numerical scheme that applies the Furlanetto \etal (2004) analytical model
to the linear density fluctuations that are the initial conditions of the
simulation.}
\label{fig:Zahn}
\end{figure}

The idea is to use the resonance associated with the hyperfine
splitting in the ground state of hydrogen (see
Section~\ref{sec:21cmAtomic}). While the CMB spectrum peaks at a
wavelength of 2 mm, it provides a still-measurable intensity at meter
wavelengths that can be used as the bright background source against
which we can see the expected 1\% absorption by neutral hydrogen along
the line of sight (Hogan \& Rees 1979; Scott \& Rees 1990). The
hydrogen gas produces 21-cm absorption if its spin temperature is
colder than the CMB and excess emission if it is hotter. Since the CMB
covers the entire sky, a complete three-dimensional map of neutral
hydrogen can in principle be made from the sky position of each
absorbing gas cloud together with its redshift $z$.

Because the smallest angular size resolvable by a telescope is proportional
to the observed wavelength, radio astronomy at wavelengths as large as a
meter has remained relatively undeveloped. Producing resolved images even
of large sources such as cosmological ionized bubbles requires telescopes
which have a kilometer scale. It is much more cost-effective to use a large
array of thousands of simple antennas distributed over several kilometers,
and to use computers to cross-correlate the measurements of the individual
antennas and combine them effectively into a single large telescope. The
new experiments are being placed mostly in remote sites, because the cosmic
wavelength region overlaps with more mundane terrestrial
telecommunications.

In approaching redshifted 21-cm observations, although the first inkling
might be to consider the mean emission signal, the signal is orders of
magnitude fainter than foreground synchrotron emission from relativistic
electrons in the magnetic field of our own Milky Way (Furlanetto \etal
2006) as well as other galaxies (Di Matteo \etal 2002). Thus cosmologists
have focused on the expected characteristic variations in $T_b$, both with
position on the sky and especially with frequency, which signifies redshift
for the cosmic signal. The synchrotron foreground is expected to have a
smooth frequency spectrum, and so it is possible to isolate the
cosmological signal by taking the difference in the sky brightness
fluctuations at slightly different frequencies (as long as the frequency
separation corresponds to the characteristic size of ionized bubbles). The
21-cm brightness temperature depends on the density of neutral hydrogen. As
explained in the previous subsection, large-scale patterns in the
reionization are driven by spatial variations in the abundance of galaxies;
the 21-cm fluctuations reach $\sim$5 mK (root mean square) in brightness
temperature (Figure~\ref{fig:Mellema}) on a scale of 10 Mpc
(comoving). While detailed maps will be difficult to extract due to the
foreground emission, a statistical detection of these fluctuations (through
the power spectrum) is expected to be well within the capabilities of the
first-generation experiments now being built (Bowman \etal 2006; McQuinn
\etal 2006). Current work suggests that the key information on the topology
and timing of reionization can be extracted statistically.

\begin{figure}
\centering
\includegraphics[width=5in]{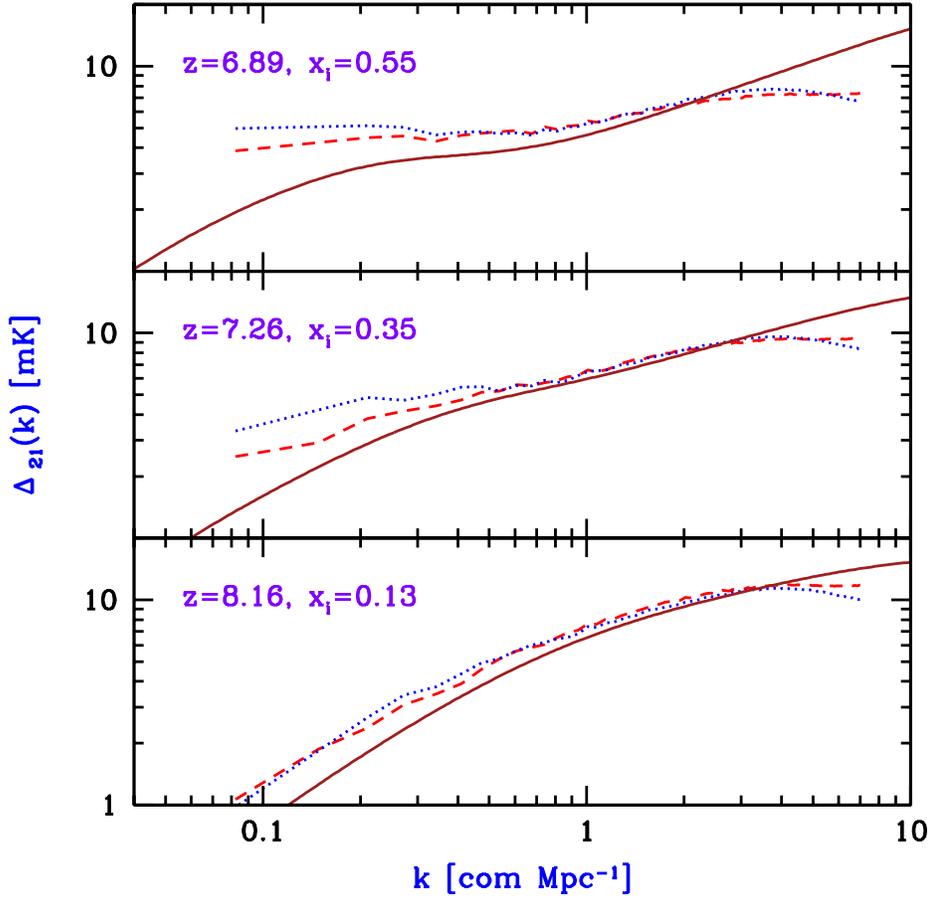}
\caption{21-cm power spectrum, comparing the analytical model of Barkana
(2007) (solid curves) to those from the simulation of Zahn \etal (2006)
(dashed curves) and from their numerical scheme based on the final ionizing
sources (dotted curves). The results are shown at several different
redshifts, as indicated in each panel. At each redshift $z$, the value of
the efficiency in the analytical model is adjusted in order to match the
mean global ionized fraction $x_i$ from the simulation. While the model and
the numerical scheme include some approximations and the simulation has
resolution and size limitations, they all show the same overall trend: as
reionization advances, the power spectrum flattens, with the increased
power on large scales ($k \sim 0.1$ Mpc$^{-1}$) reflecting the increasing
size of bubbles that are due to clustered groups of galaxies.}
\label{fig:Analytic}
\end{figure}

While numerical simulations of reionization are now reaching the
cosmological box sizes needed to predict the large-scale topology of the
ionized bubbles, they do this at the price of limited small-scale
resolution. These simulations cannot yet follow in any detail the formation
of individual stars within galaxies, or the feedback that stars produce on
the surrounding gas, such as photoheating or the hydrodynamic and chemical
impact of supernovae, which blow hot bubbles of gas enriched with the
chemical products of stellar nucleosynthesis. Thus, the simulations cannot
directly predict whether the stars that form during reionization are
similar to the stars in the Milky Way and nearby galaxies or to the
primordial $100 M_{\odot}$ behemoths. They also cannot determine whether
feedback prevents low-mass dark matter halos from forming stars. Thus,
models are needed that make it possible to vary all these astrophysical
parameters of the ionizing sources and to study the effect on the 21-cm
observations. For example, Furlanetto \etal (2004) developed an analytical
model that allows the calculation of the one-point bubble distribution,
i.e., the probability distribution (at a given redshift) of the size of the
ionizing bubble surrounding a random point in space. Zahn \etal (2006) have
considered numerical schemes that apply the Furlanetto \etal (2004) model
to either the initial conditions of their simulation or to part of its
results (Figure~\ref{fig:Zahn}). Barkana (2007) has generalized the
Furlanetto \etal (2004) model and presented an analytical model that
calculates the 21-cm two-point correlation (or equivalently, the power
spectrum), a quantity of interest for the upcoming 21-cm observations
(Figure~\ref{fig:Analytic}).

The theoretical expectations presented here for reionization and for the
21-cm signal are based on rather large extrapolations from observed
galaxies to deduce the properties of much smaller galaxies that formed at
an earlier cosmic epoch. Considerable surprises are thus possible, such as
an early population of quasars or even unstable exotic particles that
emitted ionizing radiation as they decayed. In any case, the forthcoming
observational data in 21-cm cosmology should make the next few years a very
exciting time.

\subsection{Multiple uses in the era before reionization}

\begin{figure}
\centering
\includegraphics[width=5in]{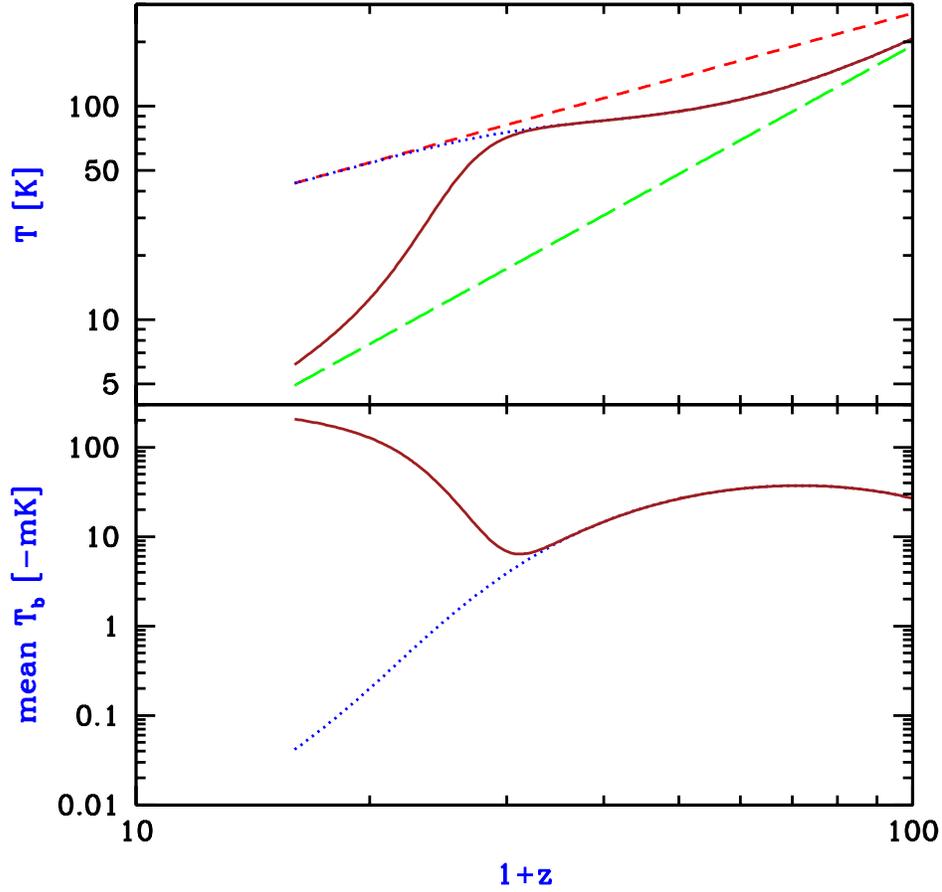}
\caption{Redshift evolution of various mean temperatures (from Barkana \&
Loeb 2005b). The mean 21-cm spin temperature is shown in the upper panel
for adiabatic IGM gas with scattering of stellar \Lya photons (solid curve)
and without it (dotted curve). Also shown for comparison are the gas
temperature (long-dashed curve) and the CMB temperature (short-dashed
curve). In adiabatic expansion, the CMB temperature cools approximately as
$\propto (1+z)$ and the gas cools faster, $\propto (1+z)^2$. The mean 21-cm
brightness temperature offset relative to the CMB is shown in the lower
panel, with \Lya scattering (solid curve) and without it (dotted
curve). The \Lya flux is calculated assuming that galaxies with Pop III
stars form in dark matter halos where the gas cools efficiently via atomic
cooling. The star formation efficiency is normalized so that \Lya
scattering becomes significant (i.e., brings the value of $x_{\rm tot}$
from section~\ref{sec:21cmAtomic} to unity) at redshift 20.}
\label{Tevol}
\end{figure}

A detection of the cosmological 21-cm signal will open a new window on
the universe and likely motivate a second generation of more powerful
telescopes. These will be used to obtain three-dimensional maps of
atomic hydrogen during reionization as well as statistical
power-spectrum measurements at even higher redshifts (using
wavelengths at which the foregrounds are brighter and thus more
difficult to remove). Since the 21-cm measurements are sensitive to
any difference between the hydrogen temperature and the CMB
temperature, the potential reach of 21-cm cosmology extends down to a
cosmic age of $\sim$6 million years $(z \sim 200)$, when the IGM first
cooled below the CMB temperature (an event referred to as ``thermal
decoupling'') due to the cosmic expansion. At those very high
redshifts, the cosmic gas was so dense that atomic collisions kept the
21-cm spin temperature near that of the gas. As the universe expanded,
if there were no stars, CMB scattering would overcome the decreasing
collision rate, drive the spin temperature to the CMB temperature, and
eliminate the 21-cm signal at $z \lesssim 30$ (Figure~\ref{Tevol}).
Instead, the formation of a significant galaxy population by redshift
30 is expected to drive the spin temperature back toward the gas
temperature through the indirect mechanism of Ly$\alpha$ scattering
(Wouthuysen 1952; Field 1958; Madau et al.\ 1997; Chen \&
Miralda-Escud\'e 2004). Thus, it should be possible to detect a 21-cm
signal throughout this redshift range, in absorption (as long as the
gas is cooler than the CMB).

At high redshifts prior to reionization, spatial perturbations in the
thermodynamic gas properties are linear and can be predicted precisely
(see section~\ref{sec:lin}). While early collapsed mini-halos are
expected to produce a 21-cm signal (Iliev et al.\ 2003), in most
scenarios this signal is swamped by that from the IGM (Oh \& Mack
2003; Furlanetto \& Oh 2006). Thus, if the gas is probed with the
21-cm technique then it becomes a promising tool of fundamental,
precision cosmology, able to probe the primordial power spectrum of
density fluctuations imprinted in the very early universe, perhaps in
an era of cosmic inflation. The 21-cm fluctuations can be measured
down to the smallest scales where the baryon pressure suppresses gas
fluctuations, while the CMB anisotropies are damped on small scales
(through the so-called Silk damping). This difference in damping
scales can be seen by comparing the baryon-density and
photon-temperature power spectra in Figure~\ref{fig:photons}.  Since
the 21-cm technique is also three-dimensional (while the CMB yields a
single sky map), there is a much large potential number of independent
modes probed by the 21-cm signal: $N_{\rm 21-cm}\sim 3 \times 10^{16}$
compared to $N_{\rm cmb} \sim 2\times 10^7$ (Loeb \& Zaldarriaga
2004). This larger number should provide a measure of non-Gaussian
deviations to a level of $\sim N_{\rm 21 cm}^{-1/2}$, constituting a
test of the inflationary origin of the primordial inhomogeneities
which are expected to possess non-Gaussian deviations $\gtrsim
10^{-6}$.

\begin{figure}
\centering
\includegraphics[width=5in]{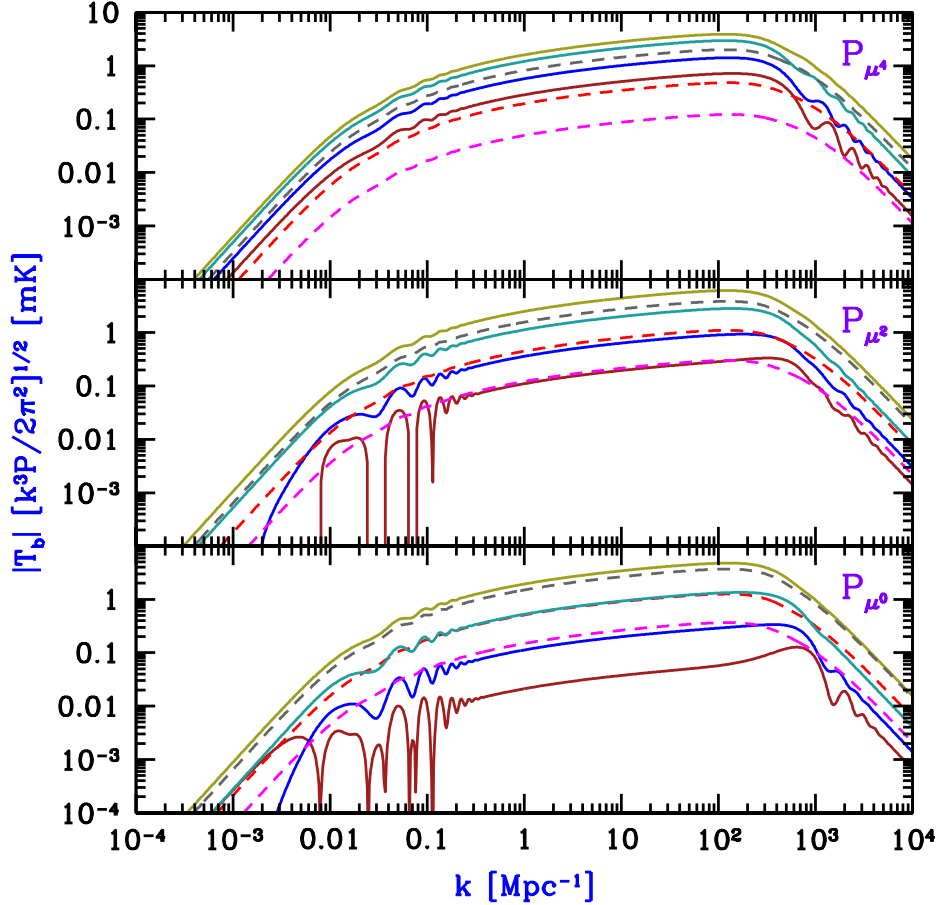}
\caption{Power spectra of 21-cm brightness fluctuations versus comoving
wavenumber (Barkana \& Loeb 2005c; Naoz \& Barkana 2005). We show the
three power spectra that are separately observable, $P_{\mu^4}$ (upper
panel), $P_{\mu^2}$ (middle panel), and $P_{\mu^0}$ (lower panel). In
each case we show redshifts 200, 150, 100, 50 (solid curves, from
bottom to top), 35, 25, and 20 (dashed curves, from top to bottom).}
\label{21cmP}
\end{figure}

An important cross-check on these measurements is possible by measuring the
particular form of anisotropy, expected in the 21-cm fluctuations, that is
caused by gas motions along the line of sight (Kaiser 1987; Bharadwaj \&
Ali 2004; Barkana \& Loeb 2005a). This anisotropy, expected in any
measurement of density that is based on a spectral resonance or on redshift
measurements, results from velocity compression. Consider a photon
traveling along the line of sight that resonates with absorbing atoms at a
particular point. In a uniform, expanding universe, the absorption optical
depth encountered by this photon probes only a narrow strip of atoms, since
the expansion of the universe makes all other atoms move with a relative
velocity that takes them outside the narrow frequency width of the
resonance line. If there is a density peak, however, near the resonating
position, the increased gravity will reduce the expansion velocities around
this point and bring more gas into the resonating velocity width. This
effect is sensitive only to the line-of-sight component of the velocity
gradient of the gas, and thus causes an observed anisotropy in the power
spectrum even when all physical causes of the fluctuations are
statistically isotropic. Barkana \& Loeb (2005a) showed that this
anisotropy is particularly important in the case of 21-cm
fluctuations. When all fluctuations are linear, the 21-cm power spectrum
takes the form (Barkana \& Loeb 2005a) \beq P_{\rm 21-cm}({\bf k}) = \mu^4
P_{\rho}(k) + 2 \mu^2 P_{\rho - {\rm iso}} (k) + P_{\rm iso}\ , \eeq where
$\mu = \cos \theta$ in terms of the angle $\theta$ between the wavevector
${\bf k}$ of a given Fourier mode and the line of sight, $P_{\rm iso}$ is
the isotropic power spectrum that would result from all sources of 21-cm
fluctuations without velocity compression, $P_{\rho}(k)$ is the 21-cm power
spectrum from gas density fluctuations alone, and $P_{\rho - {\rm iso}}
(k)$ is the Fourier transform of the cross-correlation between the density
and all sources of 21-cm fluctuations. The three power spectra can also be
denoted $P_{\mu^4}(k)$, $P_{\mu^2}(k)$, and $P_{\mu^0}(k)$, according to
the power of $\mu$ that multiplies each term. The prediction for these
power spectra at high redshift ($z > 20$), neglecting the effects of any
stellar radiation, are shown in Figure~\ref{21cmP}. At these redshifts, the
21-cm fluctuations probe the infall of the baryons into the dark matter
potential wells (Barkana \& Loeb 2005c). The power spectrum shows remnants
of the photon-baryon acoustic oscillations on large scales, and of the
baryon pressure suppression on small scales (Naoz \& Barkana 2005).

Once stellar radiation becomes significant, many processes can
contribute to the 21-cm fluctuations. The contributions include
fluctuations in gas density, temperature, ionized fraction, and \Lya
flux. These processes can be divided into two broad categories: The
first, related to {\it ``physics''}, consists of probes of
fundamental, precision cosmology, and the second, related to {\it
``astrophysics''}, consists of probes of stars. Both categories are
interesting -- the first for precision measures of cosmological
parameters and studies of processes in the early universe, and the
second for studies of the properties of the first galaxies. However,
the astrophysics depends on complex non-linear processes (collapse of
dark matter halos, star formation, supernova feedback), and must be
cleanly separated from the physics contribution, in order to allow
precision measurements of the latter.  As long as all the fluctuations
are linear, the anisotropy noted above allows precisely this
separation of the physics from the astrophysics of the 21-cm
fluctuations (Barkana \& Loeb 2005a). In particular, the
$P_{\mu^4}(k)$ is independent of the effects of stellar radiation, and
is a clean probe of the gas density fluctuations. Once non-linear
terms become important, there arises a significant mixing of the
different terms; in particular, this occurs on the scale of the
ionizing bubbles during reionization (McQuinn et al.\ 2006).

The 21-cm fluctuations are affected by fluctuations in the Ly$\alpha$
flux from stars, a result that yields an indirect method to detect and
study the early population of galaxies at $z \sim 20$ (Barkana \& Loeb
2005b). The fluctuations are caused by biased inhomogeneities in the
density of galaxies, along with Poisson fluctuations in the number of
galaxies. Observing the power-spectra of these two sources would probe
the number density of the earliest galaxies and the typical mass of
their host dark matter halos. Furthermore, the enhanced amplitude of
the 21cm fluctuations from the era of \Lya coupling improves
considerably the practical prospects for their detection. Precise
predictions account for the detailed properties of all possible
cascades of a hydrogen atom after it absorbs a photon (Hirata 2006;
Pritchard \& Furlanetto 2006a). Around the same time, X-rays may also
start to heat the cosmic gas and to indirectly generate Ly$\alpha$
photons, producing strong 21-cm fluctuations due to fluctuations in
the X-ray flux (Chuzhoy et al.\ 2006, Pritchard \& Furlanetto 2006b).

\section{Future prospects}

Understandably, astronomers are eager to start tuning into the cosmic radio
channels of 21-cm cosmology. The main challenge involves the Galactic
synchrotron foreground which is several orders of magnitude brighter than
the cosmic signal (for details, see Furlanetto \etal 2006). Removal of the
bright foreground is made feasible by the fact that it does not change for
slight shifts in observed wavelength while the cosmological signal does
(because different wavelengths correspond to different slices through the
``swiss cheese'' structure of the cosmic hydrogen). Primordial hydrogen
produced 21-cm absorption against the CMB sky within the redshift interval
$30 \la z\la 200$. Density fluctuations in the gas produced variable
absorption, and corresponding 21-cm brightness fluctuations during that
epoch. Probing these fluctuations in three dimensions down to small scales
would offer the biggest data set about the initial conditions of the
universe, with orders of magnitude more bits of information than the CMB
anisotropies or galaxy surveys could ever provide. Such information has the
potential to bring new clues about small deviations from Gaussianity in the
initial density field or about new species of cosmic matter.  At redshifts
$z\la 20$, the cosmic gas is expected to be heated above the CMB
temperature by radiation from the first galaxies, and if so its 21-cm
brightness would exceed that of the CMB.  At this stage, the 21-cm
fluctuations are sourced by the first galaxies which produce
ionization fraction fluctuations, Ly$\alpha$ intensity fluctuations,
temperature fluctuations, or direct 21-cm emission from mini-halos.

\begin{figure}
\centering
\includegraphics[width=4in]{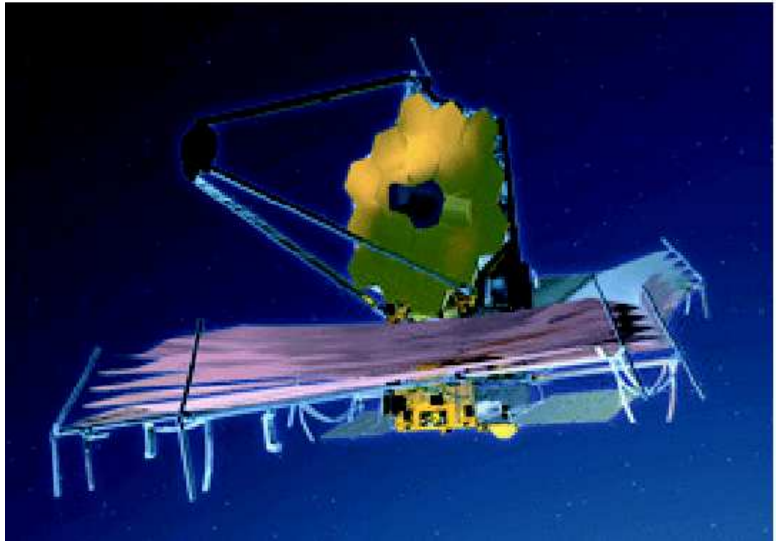}
\caption{A sketch of the current design for the {\it James Webb Space
Telescope}, the successor to the {\it Hubble Space Telescope}\/ to be
launched in 2013 (see http://www.jwst.nasa.gov/). The current design
includes a primary mirror made of beryllium which is 6.5 meters in
diameter as well as an instrument sensitivity that spans the full
range of infrared wavelengths of 0.6--28$\mu$m and will allow
detection of some of the first galaxies in the infant Universe. The
telescope will orbit 1.5 million km from Earth at the Lagrange L2
point. Note that the sun shield (the large flat screen in the image)
is 22m$\times$10m in size.}
\label{jwst}
\end{figure}

\begin{figure}
\centering
\includegraphics[width=4in]{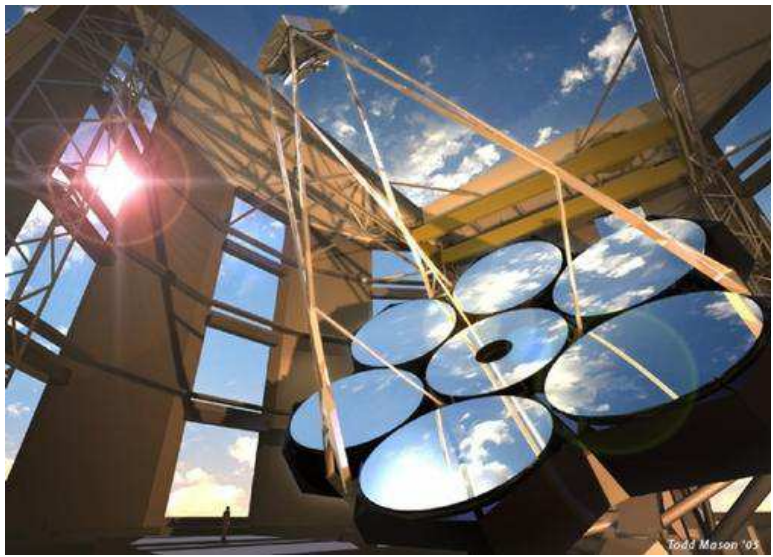}
\caption{Artist's conception of the design for one of the future giant
telescopes that could probe the first generation of galaxies from the
ground. The {\it Giant Magellan Telescope}\/ ({\it GMT}\/) will
contain seven mirrors (each 8.4 meter in diameter) and will have a
resolving power equivalent to a 24.5 meter (80 foot) primary
mirror. For more details see http://www.gmto.org/ .}
\label{gmt}
\end{figure}

In parallel to the search for redshifted 21-cm fluctuations, future
infrared telescopes will search directly for the early galaxies that
sculpted the ``swiss cheese'' topology in the primordial hydrogen.
Within the next decade, NASA plans to launch an infrared space telescope ({\it
JWST}\/; Figure~\ref{jwst}) that will image some of the earliest
sources of light (stars and black holes) in the Universe. In parallel,
there are several initiatives to construct large-aperture infrared
telescopes on the ground with the same goal in mind (see
http://www.eso.org/projects/owl/; http://celt.ucolick.org/;
http://www.gmto.org/).

The next generation of ground-based telescopes will have an effective
diameter of 20-30 meters (Figure~\ref{gmt}), and together with the
{\it JWST}\/ (which will not be affected by the atmosphere) they will
be able to image and make spectral studies of the early galaxies at
redshifts 10-30. Given that these galaxies also ionized their
environments, their locations should correlate with cavities in the
21-cm emission (Wyithe \etal 2005; Wyithe \& Loeb 2006). Within a
decade it should be possible to explore the environmental influence of
the first galaxies by using both radio and infrared instruments in
concert. The challenge is for the theorists to predict this
interaction reliably through models and numerical simulations before
it is actually observed.

\ack

The authors acknowledge support by the Israel - U.S. Binational
Science Foundation grant 2004386, the Israel Science
Foundation grant 629/05 (for RB) and Harvard University funds
(for AL).

\References

\item[] Abel T L, Bryan G L, Norman M L 2002 {\it Science} \textbf{295} 
93

\item[] Allison A C and Dalgarno A 1969 {\it Astrophys. J.} \textbf{158} 
423

\item[] Arons J and Wingert D W 1972 {\it Astrophys. J.} \textbf{177} 1

\item[] Babich D and Loeb A 2006 {\it Astrophys. J.} \textbf{640} 1

\item[] Barkana R 2006a {\it Science} \textbf{313} 931

\item[] Barkana R 2007 {\it Mon. Not. Roy. Astron. Soc.} in press 

\item[] Barkana R and Loeb A 2001 {\it Phys. Rep.} \textbf{349} 125

\item[] Barkana R and Loeb A 2004a {\it Astrophys. J.} \textbf{601} 64

\item[] Barkana R and Loeb A 2004b {\it Astrophys. J.} \textbf{609} 474

\item[] Barkana R and Loeb A 2005a {\it Astrophys. J. Lett.} 
\textbf{624} L65

\item[] Barkana R and Loeb A 2005b {\it Astrophys. J.}
\textbf{626} 1

\item[] Barkana R and Loeb A 2005c {\it Mon. Not. Roy. Astron. Soc.
Lett.}  \textbf{363} L36

\item[] Bennett C L \etal 1996 {\it Astrophys. J. Lett.} \textbf{464} L1

\item[] Benson A J, Bower, R G, Frenk C S, Lacey C G, Baugh C M and 
Cole S 2003 {\it Astrophys. J.} \textbf{599} 38

\item[] Bharadwaj S and Ali S S 2004 {\it Mon. Not. Roy. Astron. Soc.} 
\textbf{352} 142

\item[] Bowman J D, Morales M F and Hewitt J N 2006 
{\it Astrophys. J.} \textbf{638} 20

\item[] Bromm V, Coppi P S, Larson R B 2002 {\it Astrophys. J.} 
\textbf{564} 23

\item[] Bromm V and Larson R B 2004 {\it Ann. Rev. Astron. \& 
Astrophys.} \textbf{42} 79

\item[] Bromm V and Loeb A 2004 {\it New Astronomy} \textbf{9} 353

\item[] Bundy K et al.\ 2006 {\it Astrophys. J.} \textbf{651} 120 

\item[] Chen X and Miralda-Escud\'e J 2004 {\it Astrophys. J.}
\textbf{602} 1

\item[] Chuzhoy L, Alvarez M A and Shapiro P R 2006 
{\it Astrophys. J. Lett.} \textbf{648} L1

\item[] Chuzhoy L and Shapiro P R 2006 {\it Astrophys. J.}
\textbf{651} 1

\item[] Ciardi B, Ferrara A and White S D M 2003 {\it Mon. Not. Roy. 
Astron. Soc.} \textbf{344} L7

\item[] Ciardi B and Loeb A 2000 {\it Astrophys. J.} \textbf{540} 687

\item[] Cole S \etal 2005 {\it Mon. Not. R. Astron. Soc.} \textbf{362} 
505

\item[] Dijkstra M, Haiman Z, Rees M J and Weinberg D H 2004 {\it
Astrophys. J.} \textbf{601} 666

\item[] Di Matteo T, Perna R, Abel T and Rees M J 2002 
{\it Astrophys. J.} \textbf{564} 576

\item[] Efstathiou G 1992 {\it Mon. Not. Roy Astron. Soc.}  \textbf{256} 43

\item[] Eisenstein D J \etal 2005 {\it Astrophys. J.} \textbf{633} 560

\item[] Fan X \etal 2002 {\it Astron. J.} \textbf{123} 1247

\item[] Fan X \etal 2003 {\it Astron. J.} \textbf{125} 1649

\item[] Fan X \etal 2005 (available at 
http://www.arxiv.org/abs/astro-ph/0512082)

\item[] Fan X \etal 2006a {\it Astron. J.} \textbf{132} 117

\item[] Fan X, Carilli C L and Keating B 2006b {\it Ann. Rev. Astron. 
\& Astrophys.} \textbf{44} 415

\item[] Field G B 1958 {\it Proc. IRE} \textbf{46} 240

\item[] Field G B 1959a {\it Astrophys. J.} \textbf{129} 536

\item[] Field G B 1959b {\it Astrophys. J.} \textbf{129} 551

\item[] Field G B 1972 {\it Ann. Rev. Astr. \& Astrophys.} \textbf{10} 
227

\item[] Fukugita M and Kawasaki M 1994 {\it Mon. Not. Roy. Astron. Soc.} 
\textbf{269} 563

\item[] Furlanetto S R and Loeb A 2003 {\it Astrophys. J.}
\textbf{588} 18

\item[] Furlanetto S R and Oh S P 2006 {\it Astrophys. J.}
\textbf{652} 849 

\item[] Furlanetto, S R, Oh, S P and Briggs, F 2006 {\it Phys. Rep.}
in press (available at http://arxiv.org/abs/astro-ph/0608032)

\item[] Furlanetto S R, Zaldarriaga M and Hernquist L 2004 
{\it Astrophys. J.} \textbf{613} 1

\item[] Gehrels N \etal 2004 {\it Astrophys. J.} \textbf{611} 1005

\item[] Gnedin N Y 2000 {\it Astrophys. J.} \textbf{542} 535 

\item[] Gnedin N Y and Hui L 1998 {\it Mon. Not. Roy Astron. Soc.}
\textbf{296} 44 

\item[] Gnedin N Y and Ostriker J P 1997 {\it Astrophys. J.}  \textbf{486}
581

\item[] Goodman J 1995 {\it Phys. Rev. D} \textbf{52} 1821

\item[] Gunn J E, Peterson B A 1965 {\it Astrophys. J.} \textbf{142} 
1633 

\item[] Haiman Z and Loeb A 1997 {\it Astrophys. J.} \textbf{483} 21 

\item[] Haiman Z, Rees M J, Loeb A 1997 {\it Astrophys. J.} \textbf{476} 
458; erratum -- {\it Astrophys. J.} \textbf{484} 985

\item[] Haiman Z, Thoul A A and Loeb A 1996 {\it Astrophys. J.}
\textbf{464} 52

\item[] Haislip J \etal 2006 {\it Nature} \textbf{440} 181

\item[] Hirata C M 2006 {\it Mon. Not. Roy. Astron. Soc.} \textbf{367} 259 

\item[] Hogan C J and Rees M J 1979 {\it Mon. Not. Roy. Astron. Soc.} 
\textbf{188} 791

\item[] Hu E M, Cowie L L, McMahon R G, Capak P, 
Iwamuro F, Kneib J-P, Maihara T and Motohara K 2002 {\it
Astrophys. J. Lett.} \textbf{568} L75

\item[] Iliev I T, Scannapieco E, Martel H and Shapiro P R 2003 
{\it Mon. Not. Roy. Astron. Soc.} \textbf{341} 81 

\item[] Iye M \etal 2006 {\it Nature} \textbf{443} 186

\item[] Kaiser N 1984 {\it Astrophys. J. Lett.} \textbf{284} L9

\item[] Kaiser N 1987 {\it Mon. Not. Roy. Astron. Soc.} \textbf{227} 1

\item[] Kamionkowski M, Spergel D N and Sugiyama N 1994
{\it Astrophys. J. Lett.} \textbf{426} L57

\item[] Kitayama T and Ikeuchi S 2000 {\it Astrophys. J.}  \textbf{529} 615

\item[] Kolb E W and Turner M S 1990 {\it The Early Universe} (Redwood
City, CA: Addison-Wesley)

\item[] Lamb D Q and Reichart D E 2000 {\it Astrophys. J.} 
\textbf{536} 1

\item[] Lidz A, Oh S P and Furlanetto S R 2006 {\it Astrophys. J.
Lett.} \textbf{639} L47

\item[] Loeb A 2006 {\it First Light}, SAAS-Fee lecture notes, to be
published by Springer Verlag (available at
http://arxiv.org/abs/astro-ph/0603360)

\item[] Loeb A and Zaldarriaga M 2004 {\it Phys. Rev. Lett.} 
\textbf{92} 211301 

\item[] Ma C and Bertschinger E 1995 {\it Astrophys. J.} \textbf{455} 7

\item[] Madau P, Meiksin A and Rees M J 1997 {\it Astrophys. J.} 
\textbf{475} 429 

\item[] McQuinn M, Zahn O, Zaldarriaga M, Hernquist L and
Furlanetto S R 2006 {\it Astrophys. J.} in press (available at
http://www.arxiv.org/abs/astro-ph/0512263)

\item[] Meiksin A and Madau P 1993 {\it Astrophys. J.} 
\textbf{412} 24

\item[] Mellema G, Iliev I T, Pen U-L and Shapiro P R 2006 
{\it Mon. Not. Roy. Astron. Soc.} in press (available at
http://www.arxiv.org/abs/astro-ph/0603518)

\item[] Mesinger A and Haiman Z 2004 {\it Astrophys. J. Lett.}
\textbf{611} 69

\item[] Miralda-Escud\'e J 1998 {\it Astrophys. J.} \textbf{501} 15

\item[] Miralda-Escud\'e J 2000 {\it Astrophys. J. Lett.} \textbf{528} 
L1

\item[] Miralda-Escud\'e J and Ostriker J P 1990 {\it Astrophys. J.}
\textbf{350} 1

\item[] Miralda-Escud\'e J and Rees M J 1998 {\it Astrophys. J.}
\textbf{497} 21

\item[] Naoz S and Barkana R 2005 {\it Mon. Not. Roy. Astron. Soc.} 
\textbf{362} 1047

\item[] Naoz S, Noter S and Barkana R 2006 {\it
Mon. Not. Roy. Astron. Soc. Lett.} in press (available at
http://www.arxiv.org/abs/astro-ph/0604050).

\item[] Navarro J F and Steinmetz M 1997 {\it Astrophys. J.}  \textbf{478}
13 

\item[] Oh S P 2001 {\it Astrophys. J.} \textbf{553} 499

\item[] Oh S P and Mack K J 2003 {\it Mon. Not. Roy. Astron. Soc.} 
\textbf{346} 871 

\item[] Peebles P J E 1980 {\it The Large-Scale Structure of the 
Universe} (Princeton: Princeton University Press) 

\item[] Peebles P J E 1984 {\it Astrophys. J.} \textbf{277} 470

\item[] Peebles P J E 1993 {\it Principles of Physical Cosmology}
(Princeton: Princeton University Press) 

\item[] Peebles P J E and Yu J T 1970 {\it Astrophys. J.} \textbf{162} 
815

\item[] Pritchard J R and Furlanetto S R 2006a {\it Mon. Not. Roy. 
Astron. Soc.} \textbf{367} 1057

\item[] Pritchard J R and Furlanetto S R 2006b {\it Mon. Not. Roy. 
Astron. Soc.} submitted, astro-ph/0607234

\item[] Purcell E M and Field G B 1956 {\it Astrophys. J.} \textbf{124} 
542

\item[] Quinn T, Katz N and Efstathiou G 1996 {\it
Mon. Not. Roy Astron. Soc. Lett.} \textbf{278} 49

\item[] Rees M J 1986 {\it Mon. Not. Roy. Astron. Soc.} \textbf{222} 27

\item[] Rees M J and Sciama D W 1968 {\it Nature} \textbf{217} 511

\item[] Sachs R K and Wolfe A M 1967 {\it Astrophys. J.} \textbf{147} 
73

\item[] Scott D and Rees M J 1990 {\it Mon. Not. Roy. Astron. Soc.} 
\textbf{247} 510

\item[] Seljak U and Zaldarriaga M 1996 {\it Astrophys. J.} 
\textbf{469} 437

\item[] Shapiro P R, Giroux M L 1987 {\it Astrophys. J. Lett.} 
\textbf{321} L107

\item[] Shapiro P R, Giroux M L and Babul A 1994, {\it Astrophys. J.} 
\textbf{427} 25

\item[] Silk J 1968 {\it Astrophys. J.} \textbf{151} 459 

\item[] Spergel D N \etal 2006 {\it Astrophys. J.} in press (available 
at http://www.arxiv.org/abs/astro-ph/0603449)

\item[] Sunyaev R A and Zeldovich Y B 1970 {\it APSS} \textbf{7} 3

\item[] Tegmark M \etal 1997 {\it Astrophys. J.} \textbf{474} 1

\item[] Tegmark M, Silk J and Blanchard A 1994 {\it Astrophys. J.} 
\textbf{420} 484 

\item[] Thoul A A and Weinberg D H 1996 {\it Astrophys. J.}
\textbf{465} 608 

\item[] Totani T, Kawai N, Kosugi G, Aoki K, Yamada T, Iye M, Ohta K 
and Hattori T 2006 {\it Pub. Astron. Soc. Japan} \textbf{58} 485

\item[] Viel M, Haehnelt M G and Lewis A 2006 {\it
Mon. Not. Roy. Astron. Soc. Lett.}  \textbf{370} L51

\item[] Weinberg D H, Hernquist L and Katz N 1997 {\it Astrophys. J.}
\textbf{477} 8

\item[] Weinberg S 1972 {\it Gravitation and Cosmology} (New York:
Wiley)

\item[] White R L, Becker R H, Fan X, Strauss M A 2003 {\it Astron. J.} 
\textbf{126} 1

\item[] Wouthuysen S A 1952 {\it Astron. J.} \textbf{57} 31 

\item[] Wu K K S, Lahav O and Rees M J 1999 {\it Nature} \textbf{397}
225

\item[] Wyithe J S B and Loeb A 2004a {\it Nature} \textbf{427} 815

\item[] Wyithe J S B and Loeb A 2004b {\it Nature} \textbf{432} 194

\item[] Wyithe J S B and Loeb A 2006 {\it Mon. Not.  Roy. Astron. Soc}
submitted (available at http://arxiv.org/abs/astro-ph/0609734)

\item[] Wyithe J S B, Loeb A and Barnes D G 2005 {\it Astrophys. J.} 
\textbf{634} 715 

\item[] Yamamoto K, Sugiyama N and Sato H 1997 {\it Phys. Rev. D} 
\textbf{56} 7566
 
\item[] Yamamoto K, Sugiyama N and Sato H 1998 {\it Astrophys. J.} 
\textbf{501} 442

\item[] Yoshida N, Omukai K, Hernquist L, and Abel T 2006 
{\it Astrophys. J.} in press (available at
http://www.arxiv.org/abs/astro-ph/0606106).

\item[] Zahn O, Lidz A, McQuinn M, Dutta S, Hernquist L, 
Zaldarriaga M and Furlanetto S R 2006 {\it Astrophys. J.} in press
(available at http://www.arxiv.org/abs/astro-ph/0604177)

\item[] Zhang W, Woosley S and MacFadyen A I 2003, {\it Astrophys. J.}
\textbf{586} 356

\item[] Zygelman B 2005 {\it Astrophys. J.} \textbf{622} 1356

\endrefs

\end{document}